\documentclass[journal=jacsat,manuscript=article]{achemso}

\usepackage[version=3]{mhchem} % Formula subscripts using \ce{}
\setcitestyle{square}
\usepackage{amssymb}
\newcommand{\braket}[2]{\langle #1|#2\rangle}

\newcommand{\Oi}{\mathbf{\Omega}_{\text{i}}}
\newcommand{\Of}{\mathbf{\Omega}_{\text{f}}}

\newcommand{\Jd}{\mathbf{J}^{\intercal}}
\newcommand{\B}{\mathbf{B}}

\newcommand{\Dd}{\mathbf{D}^{\intercal}}

\newcommand{\wf}{\omega^{f}}
\newcommand{\wi}{\omega^{i}}

\author{Enrico Tapavicza}
\affiliation[Unknown University]
{Department of Chemistry and Biochemistry, California State University, Long Beach,  1250 Bellflower Boulevard, Long Beach, CA 90840}
\email{enrico.tapavicza@csulb.edu}

\title{Conformationally controlled photochemistry studied by trajectory surface hopping}

\begin{document}

\section{Introduction}\label{sec1.1}
The influence of the conformation of flexible molecules on the photochemical reactivity has been an active field of research since the early work of Havinga \cite{Havinga1961}, which was motivated by understanding vitamin D photochemistry.
Later, Havinga formulated the concept of nonequilibrium excited state rotamers (NEER) on the basis of the reactivity of simple hexatriene \cite{Havinga1973b,Vroegop1973} molecules and vitamin D related model compounds \cite{Boomsma1975}. 
The NEER principle states that the photochemical reactivity and photoproduct distribution of a flexible molecule is dictated by the conformational equilibrium of the molecule in the electronic ground state and not by the conformational equilibrium that would result from the potential energy surface of the excited singlet state. The rationale of this argument is that after photoexcitation of the ensemble of conformers, the lifetime of the reactive excited state is too short to allow for structural adaptation to the new equilibrium of rotamers in the excited state and thus the outcome of a fast photochemical reaction is a consequence of the conformational equilibrium at the time of photoexcitation.

This principle has been found to be useful to understand the photoreactivity in a variety of hexatriene systems, most prominently in vitamin D photochemistry, but also in the photochemical synthesis of ikarugamycin \cite{Whitesell1989}, and the photochemistry of alkyl-substituted hexatrienes \cite{baldwin1969stereoselective,Jacobs1979}.
Besides hexatriene systems, conformationally controlled photochemistry has also been found involved in the reactivity of substituted ketones \cite{engel1972effect,lewis1974conformational,wagner1981kinetically,lipson1994conformational}, in photochemical reactions of bifunctional molecules \cite{wagner1983conformational}, in the triplet reactivity of conjugated dienes \cite{liu1965mechanisms}, in the photoinduced ring-opening of substituted hexadienes \cite{spangler1972influence}, in the photoreactivity of alkyl azides \cite{moriarty1970direct}, in intramolecular photocyclization of dienes \cite{oppolzer1982intramolecular}, $\alpha$-(aminoalkyl)styrenes \cite{lewis1993intramolecular} and diarylethenes \cite{matsuda2003very}, in the wavelength-dependent photochemistry of cyclocarbenes \cite{zuev2006conformational}, and the photoreactivity of bilirubin \cite{madea2020wavelength,janos2020conformational}.
With the recent developments in photochemical switches, light-driven motors \cite{browne2010making,grathwol2019azologization}, and photoactive proteins and peptides \cite{torner2021conformational}, strategies to control the conformationally dependent photochemistry has attracted new interest \cite{matsuda2003very,browne2009light,renth2013enhanced,ghosh2021dual}.
However, despite its success in predicting the outcome of numerous photochemical reactions, the NEER principle has failed in several cases \cite{flom1986dynamic,brearley1986dynamic,arai1989highly,arai1992effects,bartocci2009adiabatic,cesaretti2018photoinduced}; this anti-NEER behavior is attributed to low energetic barriers for the rotational isomerization in the excited state, allowing for rapid interconversion of rotamers in the excited states \cite{C6CP08311K}.
Furthermore, admixtures between these two extreme cases occur, where the excited state relaxation affects the distribution of rotamers to some extend \cite{Cisneros2017}.

The development of accurate quantum chemistry methods to describe electronically excited states such as  time-dependent density functional theory (TDDFT) \cite{Runge84}, complete active space self-consistent field (CASSCF) \cite{lischka2018multireference} and its perturbative variant CASPT2 \cite{andersson1990second,andersson1992second}, coupled cluster (CC) methods \cite{sneskov2012excited} allows to support photochemical studies by theoretical calculations. In a large number of early theoretical studies, electronic structure calculations based on single geometries at crucial points of the potential energy surface (PES), or minimum energy pathways (MEP) along a reaction coordinate are used to support experimental results \cite{Celani1996,Garavelli2001}. These static calculations do not include dynamic effects arising from the relaxation from the Franck-Condon region and from the electron-nuclear coupling. These effects are, however, often crucial to accurately describe the outcome of photochemical reactions. Therefore, methods that are able to describe the dynamics of the system are more promising to yield accurate results. In principle the description of photochemical dynamics requires a full quantum description of the coupled nuclear-electronic motion according to the time-dependent Schr\"odinger equation. This goal is usually out of reach for large and medium-sized molecules.
While a number of highly accurate methods, such as the multiconfigurational time-dependent Hartree \cite{beck2000multiconfiguration}, allow to carry out full-quantum dynamics simulations, these methods are limited to small systems due to their large computational cost. In addition, these methods often require precalculated potential energy surfaces \cite{curchod2018ab}, which is cumbersome for systems with a large number of degrees of freedom.
Computationally more efficient are approximative semiclassical "on-the-fly" non-adiabatic {\it ab initio} molecular dynamics (AIMD) methods \cite{Tapavicza2007,Doltsinis2002,Craig2005}, which are based on a classical description of the nuclear motion, but treat electrons quantum mechanically. Semiclassical non-adiabatic methods can be divided into i) mean-field or Ehrenfest approaches \cite{li2005ab,Tavernelli2005,andrade2009modified}, which are based on single trajectories computed on potential energy surfaces averaged over several electronic states, and ii)  trajectory surface hopping (TSH) \cite{Tully1990,zhu2001new,Doltsinis2002,Craig2005,Tapavicza2007} methods, in which a large number of trajectories is used to describe the dynamics of the ensemble.
Besides the purely semiclassical methods, the ab initio multiple spawning (AIMS) \cite{ben2000ab,Levine20083} method incorporates the quantum nature of nuclear motion to some extend.
These methods have allowed to study the dynamics of photochemical reaction mechanisms in a large number of systems.
On one hand, non-adiabatic simulations have been shown to be useful to predict excited state lifetimes, product distributions, branching ratios, and wavelength dependent product quantum yields \cite{Thompson2018,Tapavicza2018}.
On the other hand, the simulation of time-resolved spectra \cite{Wolf2014,Schalk2016,liu2020excited} and other observables has served as a means to verify time constants and branching ratios obtained from non-adiabatic simulations by direct comparison of their experimental counterparts.
In contrast to mean-field methods, in which nuclear trajectories are propagated on an excited state potential energy surface that is obtained as a superposition of several electronic states, the TSH molecular dynamics method has been shown to be particularly useful for the description of branching of the nuclear trajectories. Due to its computational efficiency, TDDFT is often the method of choice for solving the underlying electronic structure problem for electronically excited states.
In this article I will review the efforts of applying TSH methods based on TDDFT in investigating conformationally controlled photochemical reactions in my group.
I will first lay out the theoretical methods to study conformationally controlled photochemical reactions and then show how to apply these methods on the example of some systems, that have been studied in my group.

Theoretical steps include the generation of Boltzmann ensembles of structures, the computation of electronic absorption spectra of conformationally flexible molecules, the TDDFT trajectory surface hopping method (TDDFT-SH). Furthermore, I will outline how to compute wavelength dependent quantum yields using the TDDFT-SH method.

\vfill

\section{Theoretical Methods}\label{sec1.2}
\subsection{Generating Boltzmann ensembles}\label{boltzmann_ensembles}
The accurate simulation of a photochemical experiment critically depends on a reliable description of the initial conditions of the system at the time the system is photoexcited. Typically the most important parameters of interest are the nuclear positions and their momenta, but in certain cases molecular orientation and impact energies are required \cite{barbatti2011nonadiabatic}. Thus the first step is usually to generate an ensemble of structures and velocities for the TSH simulations.

Generating Boltzmann ensembles of structures of molecules for TSH is typically done by either generating a Wigner distribution \cite{bonacic2005theoretical,barbatti2010uv} of structures based on
vibrational spectra and vibrational normal modes computed within the harmonic oscillator approximation, 
or based on semiclassical molecular dynamics (MD) simulations. The Wigner approach has the advantage that it takes into account the quantum nature of the nuclei. This is particularly important if the molecule is assumed to be in the lowest vibrational states at the time of photoexcitation, which is the case for reactions at low temperatures or for reactions of rigid molecules with high vibrational frequencies compared to the available thermal energy.
In contrast, for conformationally flexible systems that possess low vibrational frequencies, one can assume a classical behavior of the nuclei and thus, a classical MD simulation should be able to accurately generate a Boltzmann ensemble of structures and velocities. Furthermore, a MD simulation gives access to the distribution of distinct conformations, which is not possible in the case of the Wigner distribution, because it is usually based on a single minimum energy structure.

Several approaches exist that intend to incorporate some of the advantages of classical sampling but still contain some of the quantum behavior of the nuclei. Here to mention are the approaches using ring-polymer dynamics \cite{craig2004quantum}, the
generalized Langevin equations \cite{ceriotti2009langevin}, which has been used to generate initial conditions in TSH \cite{janos2020conformational}, and mixed Franck-Condon methods combined with MD simulations, which have been used to simulate electronic spectra \cite{zuehlsdorff2018combining}.

For conformationally flexible molecules with increasing number of rotational degrees of freedom, however, sampling the ensemble of rotational isomers becomes increasingly challenging. This is especially the case when energetic minima are separated by energetic barriers that are large compared to the thermal energy available by the system. Thus to describe conformationally controlled photochemistry, special care has to be taken to ensure proper sampling of the initial ensemble of structures. To achieve this goal a variety of enhanced sampling methods \cite{frenkel2001understanding} are available, including umbrella sampling, meta-dynamics \cite{micheletti2004reconstructing}, or parallel tempering, also known as replica exchange molecular dynamics (REMD) \cite{Sugita1999}. While all these methods have been developed to accelerate sampling in systems with large energetic barriers, I will focus here on the description of REMD, which can easily be implemented in the context of classical MD and AIMD. REMD has been successfully applied to to generate Boltzmann ensembles of structures for spectra simulations and initial conditions for TSH to study conformationally flexible systems \cite{Tapavicza2011,Cisneros2017,Tapavicza2018,Thompson2018,de2018nitrogen,sofferman2021probing}.

In REMD, several molecular dynamics trajectories at different temperatures are simulated in parallel. To ensure proper sampling of a canonical ensemble, the temperature in each simulation is controlled by a Nos\'e-Hoover thermostat \cite{Nose1984}. After a certain number of MD steps, the probability that two structures from trajectories at temperature $T_i$ and $T_j$ will be switched is evaluated as
\begin{equation}
    P(i\rightarrow j)=\min\Bigg[1, \frac{\exp\Big(- \frac{E_j}{k_BT_i}-\frac{E_i}{k_BT_j}\Big)}{\exp\Big(- \frac{E_i}{k_BT_i}-\frac{E_j}{k_BT_j}\Big)}\Bigg],
\end{equation}
where $k_B$ is the Boltzmann constant and $E_i$ and $E_j$ are the total energies of structures at $T_i$ and $T_j$, respectively.
If the switching probability $P(i\rightarrow j)$ is 1, the two structures at different temperatures are switched and the nuclear coordinates are propagated at the new temperature. In case the probability is below 1, a random number from the interval [0,1] is generated and the structures are only switched if the random number is smaller than the switching probability. After a switch between two structures, velocities are rescaled to be consistent with the new temperatures. An ensemble of structures generated in this fashion has been shown to lead to the thermodynamically correct distribution of structures. Trajectories at high temperatures enable the system to overcome energetic barriers, exploring parts of the conformational space that would not have been accessible at low temperature within a reasonable simulation time. Furthermore, the method retains the relative populations of energetic minima if the simulation is carried out for a long enough time interval.

\subsection{Calculation of absorption spectra}
The initial event of a photochemical reaction is light-absorption. Therefore, a critical component in describing photochemical reactions is an accurate assessment of the light-absorption properties of the system of interest.
For instance, the prediction of wavelength dependent product quantum yields relies on an accurate prediction of electronic absorption spectra.

\begin{figure}
    \centering
    \includegraphics[scale=0.3]{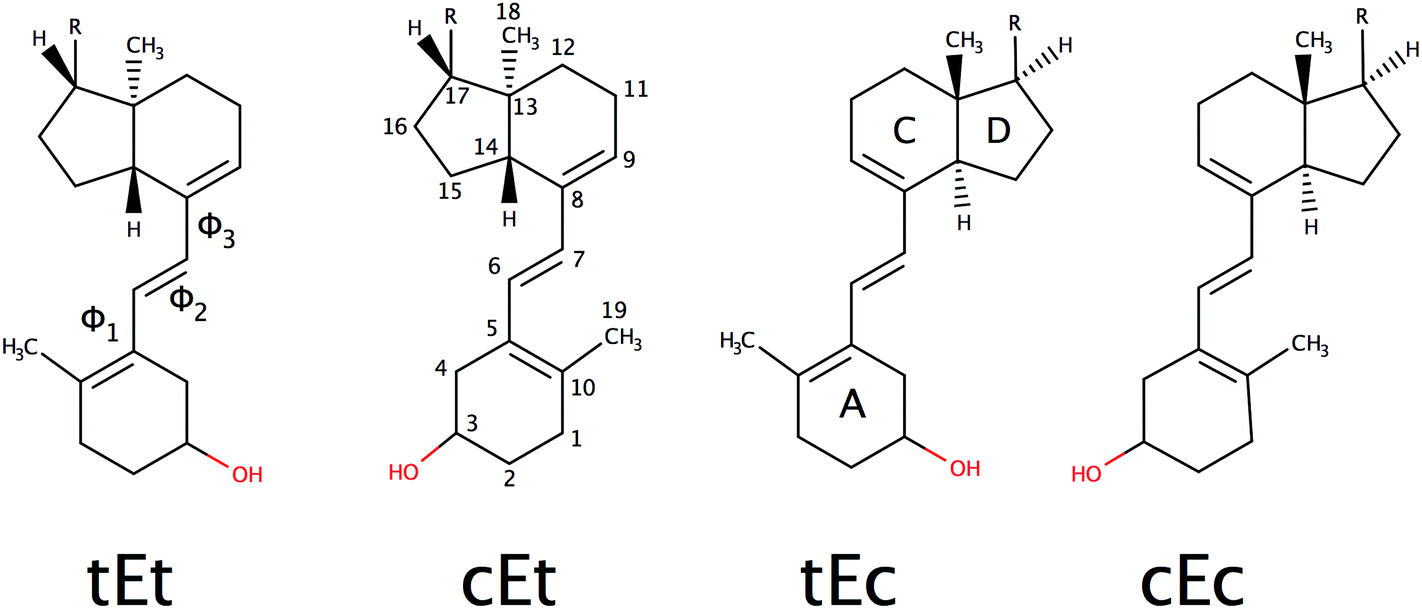}
    %\centerline{\psfig{file=c6cp08064b-f2_hi-res.eps,width=4cm}}
    \caption{Rotational isomers of truncated tachysterol. R indicates a methyl group. Dihedral angles $\Phi_1$ and $\Phi_3$ are defined by atoms C4-C5-C6-C7 and C6-C7-C8-C9, respectively. Reproduced from with permission from C. Cisneros, T. Thompson, N. Baluyot, A. C. Smith, E. Tapavicza, Phys. Chem. Chem. Phys. 19 (8), 5763-5777 (2017) the PCCP Owner Societies.}
    \label{fig:tachyrotamers}
\end{figure}

\begin{figure}
    \centering
    \includegraphics[scale=0.25]{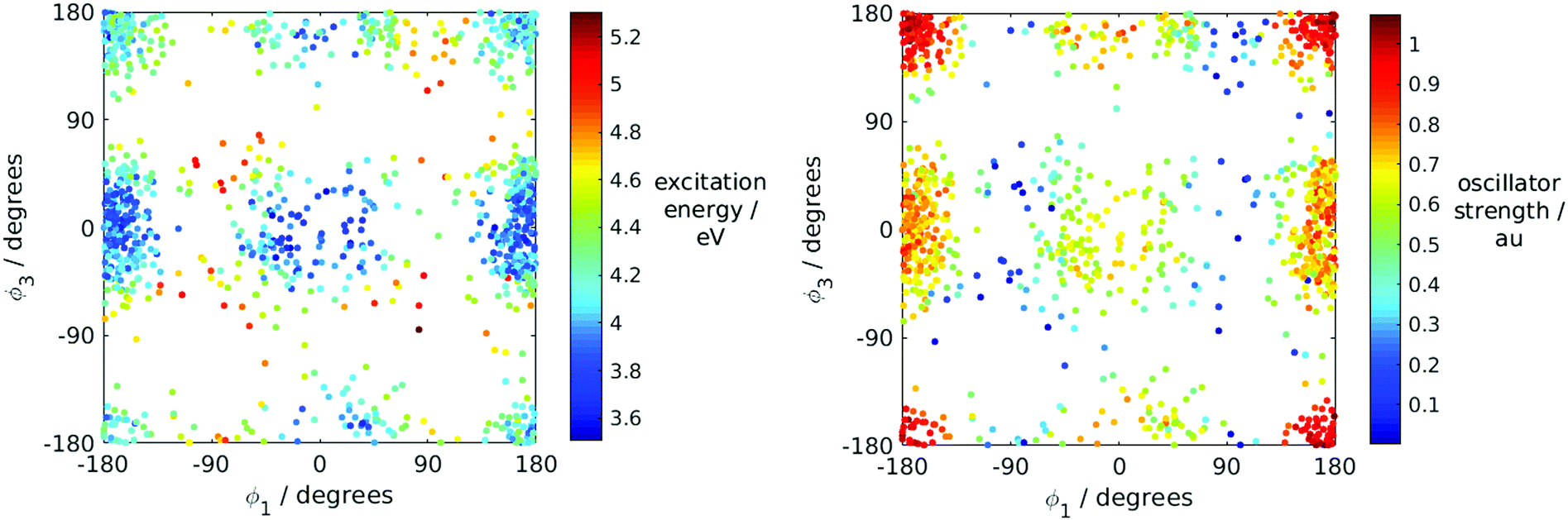}
    \caption{S$_1$ excitation energies (left) and oscillator strengths (right) as a function of dihedral angles $\Phi_1$ and $\Phi_3$, as are indicated in Figure \ref{fig:tachyrotamers}. Reproduced from C. Cisneros, T. Thompson, N. Baluyot, A. C. Smith, E. Tapavicza, Phys. Chem. Chem. Phys. 19 (8), 5763-5777 (2017) with permission from the PCCP Owner Societies.}
    \label{fig:tachyosc}
\end{figure}

\begin{figure}
    \centering
    \includegraphics[scale=0.3]{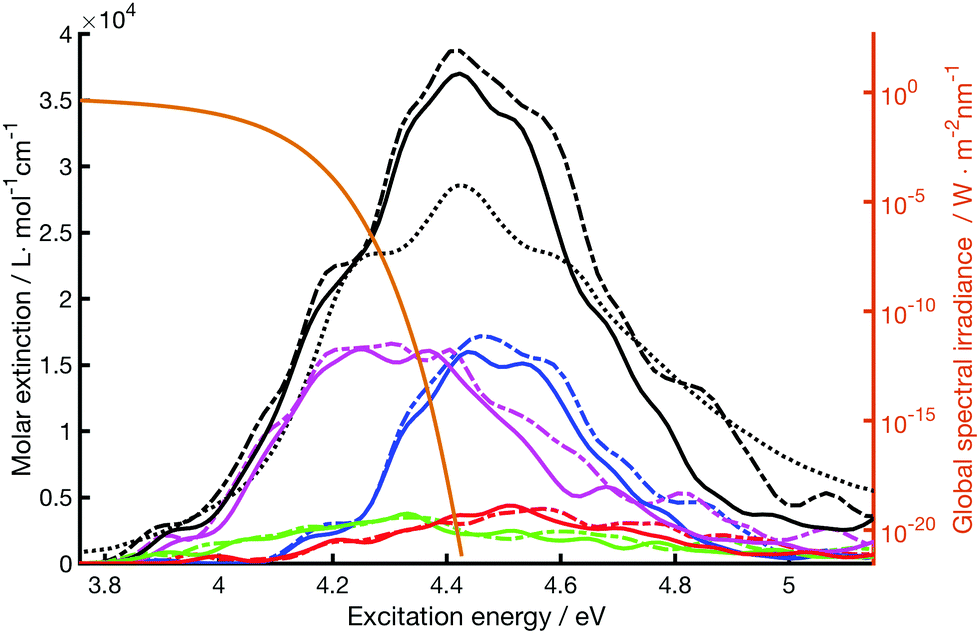}
    \caption{Tachysterol absorption spectrum for the different rotamers computed by TDPBE0 (solid) and CC2 (dashed). tEt: blue, tEc: magenta, cEt: red, cEc: green. Experimental spectrum (dotted) measured in ether.\protect\cite{MacLaughlin1982} TDPBE0 and CC2 absorption bands were shifted by to match the position of the peak maximum of the experimental spectrum. As example, the global irradiation energy density for a zenith angle of 30$^{\circ}$ is shown (brown).\protect\cite{green1974middle} These conditions approximately correspond to the irradiation in Berlin on July 8 at noon.\protect\cite{solar_position} 
Reproduced from Cisneros, C., Thompson, T., Baluyot, N., Smith, A. C. and Tapavicza, E. (2017), Phys. Chem. Chem. Phys. 19, pp. 5763–5777 with permission from the PCCP Owner Societies.}
    \label{fig:tachyspectra}
\end{figure}

Neglecting vibrational effects, for monochromatic radiation of wavelength $\lambda$,
the number of photons absorbed during a transition between two electronic states is proportional to the oscillator strength \cite{hilborn1982einstein}.
The calculation of excitation energies and oscillator strengths is routinely available from many different electronic structure methods. In wavefunction based-methods the oscillator strengths $f$ are obtained from the transition dipole moment computed for the initial and final electronic states, whereas in TDDFT $f$ is computed from the transition density between ground and excited state \cite{Furche2002}.
Recently, also machine learning models have been developed and applied to predict excitation energies and oscillator strengths \cite{Ramakrishnan2015,xue2020machine,westermayr2020machine,tapavicza2021elucidating}.
Oscillator strengths can be converted through a line function $\rho(\lambda)$ to the absorption cross section $\sigma(\lambda)$ or, equivalently, to the extinction coefficient $\epsilon(\lambda)$ \cite{hilborn1982einstein}:
\begin{equation}\label{extinction}
    \epsilon(\lambda)=2.303 \frac{4\pi^2q_e}{3\hbar c}\rho(\lambda)N_{A}f,
\end{equation}
where $\hbar$ is Planck's reduced constant, $c$ denotes the speed of light, $q_e$ is the electronic charge, and $N_A$ is Avogadro's number. Typically, Lorentzian and Gaussian lineshapes are used for broadening in case of homogeneous and inhomogeneous broadening, respectively. 
For a more accurate description of the envelope of absorption spectra conformational sampling and the inclusion of vibronic effects
might be necessary. In case of flexible molecules that exist in a large number of different conformations, a Boltzmann ensemble of structures obtained from molecular dynamics, as described in the previous section, can be used. In this case, excitation energies and oscillator strengths computed for a subset of snapshot structures from a MD or REMD trajectory are converted into absorption spectra according to Equation \ref{extinction} \cite{Epstein2013}. An average of spectra computed for a large number of snapshot structures yields the macroscopic absorption spectrum of the ensemble of structures. The procedure is exemplified for the flexible molecule tachysterol, that occurs in four major rotamers cEc, cEt, tEc, and tEt (Figure \ref{fig:tachyrotamers}).
After generating an ensemble of structures of tachysterol using REMD \cite{Cisneros2017}, excitation energies and oscillator strengths for the electronic transition between the ground state (S$_0$) and the first excited state (S$_1$) have been computed using TDDFT and the second-order approximate coupled cluster singles and doubles (CC2) method \cite{Christiansen95}. Plotting CC2 excitation energies and oscillator strengths as a function of the dihedral angles $\Phi_1$ and $\Phi_2$ (Figure \ref{fig:tachyosc}), we see that both, absorption energy and intensity, are sensitive to the conformation of the molecule.
Converting excitation energies and oscillator strengths to extinction coefficients using a Gaussian lineshape with a full width at half maximum (FWHM) of 0.1~eV, and averaging over all snapshot structures, yields the macroscopic absorption spectrum of the system (black, Figure \ref{fig:tachyspectra}). The same procedure can be applied individually to the specific rotamers (magenta, blue, green, and red in Figure \ref{fig:tachyspectra}), which allows to gauge the influence of conformation on the overall absorption spectrum.

However, in some cases the envelope of an absorption spectrum is caused by the quantum nature of the nuclear vibrations and not by the ensemble of conformations. This is particularly the case for rigid molecules that possess high vibrational frequencies.
In this case, it is necessary to apply a quantum mechanical treatment of the nuclear vibrations, which can be done using Franck-Condon methods \cite{tapavicza2019generating}.
The vibronic absorption cross section at frequency $\omega$ and temperature $T$ is given by \cite{Tapavicza2016}
\begin{eqnarray}\label{vibronic_abs}
        \sigma_{abs}(\omega,T)&=&\frac{4\pi^2\omega}{3c}|\mu_{if}|^2 \sum_{{\bf v}_\text{i}}\sum_{{\bf v}_\text{f}}P_{{\bf v}_\text{i}}(T)|\langle \theta_{{\bf v}_\text{i}}|\theta_{\bf v_\text{f}}
        \rangle|^2\delta(\Delta E_{if} + E_{\bf v_\text{i}}-E_{\bf v_\text{f}}-\omega), \\\nonumber
\end{eqnarray}
where the electronic transition dipole moment $\mu_{if}$ is defined through the wavefunctions of initial (ground) and final (excited) electronic state, $c$ is the speed of light and the Franck-Condon factors $|\langle \theta_{{\bf v}_\text{i}}|\theta_{\bf v_\text{f}}\rangle|^2$ are defined through initial and final vibrational wavefunctions $\theta_i$ and $\theta_f$, respectively.
$\Delta E_{if}$ denotes the adiabatic excitation energy, and $E_{\bf v_\text{i}}$ and $E_{\bf v_\text{f}}$ are the initial and final vibrational energies, respectively. $P_{{\bf v}_\text{i}}(T)$ denotes the occupation of initial vibrational state with quantum numbers $\bf v_\text{i}$ \cite{tapavicza2019generating}, dictated by the Boltzmann distribution.
To compute the absorption spectrum according to Equation \ref{vibronic_abs}, one needs to compute Franck-Condon factors between vibrational states of the initial and final electronic states. Two general approaches in order to do so exist: time-independent (TI) \cite{Sharp1964,Stein1973,dierksen2004vibronic,Jankowiak2007,bloino2010general}  and time-dependent (TD) \cite{kubo1952thermal,heller1981frozen,mukamel1985generating,Yan1986eigenstate,Mebel:99,Stendardo:12,Baiardi:13,Bloino:16,Tapavicza2016,patoz2018fly} approaches. In TI approaches Franck-Condon factors are computed directly in the frequency domain using analytical formulas for either the harmonic oscillator approximation \cite{Wagner1959} or for the anharmonic approximations \cite{barone2005anharmonic}. TD approaches, in contrast, use a {\it time-correlation function} $G(t)$, whose Fourier transform yields the vibronic absorption spectrum, according to
\begin{eqnarray}\label{FTem_mehler}
\sigma(\omega)& = &\frac{4\pi^2\omega}{3c} |\mu_{if}|^2 \frac{1}{2\pi}
        \int_{-\infty}^{\infty} dt \exp\big[it (\Delta E_{if}+E_{\bf 0_\text{i}}-\omega)\big] G(t),
\end{eqnarray}
where $E_{\bf 0_\text{i}}$ is the zero point vibrational energy in the ground state. Equation \ref{FTem_mehler} is formulated for the special case at zero temperature, but  expressions for finite temperature have been derived as well \cite{Peng2010,begusic2021finite}.
While TI approaches are relatively simple to implement, the computational effort for their computation increases rapidly with the number of vibrational degrees of freedom $N$, as they usually require an $N$-fold loop, unless specific algorithms, such as the Beyer-Swinehart \cite{beyer1973algorithm} and the related Rabinovich-Stein algorithm \cite{Stein1973}, are employed that lead to improved scaling with $N$. However, TD approaches exhibit better scaling with system size and are therefore suited for larger systems. Furthermore, TD approaches allow to easily incorporate Duschinsky mode \cite{Duschinsky1937} mixing, which expands the vibrational modes in the excited state $\bf{Q}_f$ as a linear combination of vibrational modes in the ground state $\bf{Q}_i$ (and reverse), according to
\begin{equation}
    \bf{Q}_f = \bf{J} \bf{Q}_i + \bf{D},
\end{equation}
where, $\bf{J}$ is the Duschinsky rotation matrix and $\bf{D}$ is the geometric displacement matrix, accounting for the difference in ground and excited state structures.

The computation of the time-correlation (or generating) function can be done efficiently based on ground and excited state equilibrium structures and their vibrational frequencies and normal modes. On the basis of the latter, the time-correlation function reads

\begin{eqnarray}\label{matrixgenfunc} 
G(t) & = & 2^{\frac{N}{2}}
\left(\frac{\det({\bf S}^{-1} \Oi\Of)}{\det({\bf L})\det({\bf M})}\right)^\frac{1}{2} \exp\big[\Dd(\Of{\bf BJM}^{-1}\Jd\Of\B-\Of{\bf B}){\bf D}\big]. \\ \nonumber 
\end{eqnarray}

Matrices $\Oi$, $\Of$, $\bf S$, $\bf B$ are diagonal and depend on time $t$ and vibrational frequencies in the ground ($\omega^i_k$) and excited state ($\omega^f_k$):
$(\Omega_\text{i})_{kk}=\wi_{k}$, $(\Omega_\text{f})_{kk}=\wf_{k}$,
$S_{kk}=\sinh(i\wf_k t)$, and $B_{kk}=\tanh(i\wf_k t/2)$.
$^\intercal$ denotes the transpose of the matrices. Matrices $\bf L$ and $\bf M$ are given by ${\bf M}= \Jd\Of{\bf BJ}+\Oi$ and
${\bf L}= \Jd\Of {\bf B}^{-1}{\bf J}+\Oi$, respectively. Using Equation \ref{FTem_mehler} together with Equation \ref{matrixgenfunc} allows to compute the absorption spectrum for the zero temperature case.
Alternatively, the time-correlation function can be computed using wavepacket propagation methods \cite{heller1981frozen,Heller1973} or semiclassical methods based on trajectories \cite{beguvsic2018fly}.

\begin{figure}
    \centering
    \includegraphics[scale=1.9]{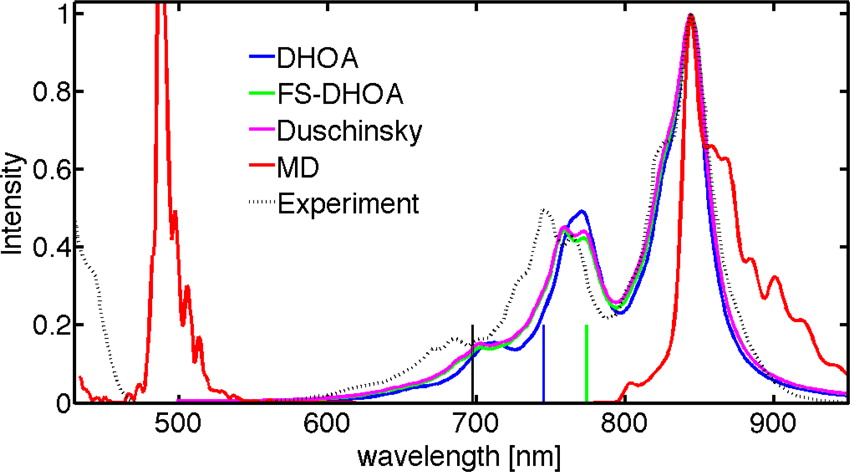}
\caption{Comparison between the calculated and experimental (dashed) spectra. The vertical excitation energy is indicated by the black bar (1.778 eV). The original position of the spectrum obtained from the displaced harmonic oscillator approximation (DHOA) is indicated by the blue bar (1.664 eV). The DHOA spectrum was red-shifted by 106 nm from the adiabatic excitation energy to match the 0–0 transition of the experimental spectrum. The frequency-shifted DHOA (FS-DHOA) spectrum (green) and the spectrum calculated using Duschinsky rotation (magenta) were shifted by 71 nm from the original position of the corrected 0–0 transition (green bar, 1.602 eV). The peak of the spectrum obtained from MD simulations (red) was shifted by 135 nm. 
Reprinted with permission from E. Tapavicza, F. Furche, D. Sundholm, J. Chem. Theory Comput. 2016, 12, 5058-5066. Copyright 2016 American Chemical Society.}
\label{importance}
\end{figure}

For rigid molecules, the previously discussed method using snapshot structures from MD trajectories can lead to an erroneous modeling of the absorption spectrum, as shown in Figure \ref{importance} for 7,7,8,8,-tetracyanoquinodimethane anion (TCNQ$^-$). As we can see from Figure \ref{importance}, the MD approach (red) is not able to reproduce the experimentally measured absorption spectrum (black, dashed), because it misses the bands between 600 and 800 nm, which originate from vibronic effects. In contrast, all approaches based on the generating function approach (DHOA, FS-DHOA, and Duschinsky) are able to account for the bands in this spectral region in TCNQ$^-$.

\subsection{Linear response time-dependent density functional surface hopping}
A mixed quantum-classical treatment of the coupled nuclear and electronic dynamics \cite{Tapavicza2013} of a photochemical reaction requires the solution  of the time-dependent electronic Schr\"odinger equation
\begin{equation}\label{Eq:tdse}
i\hbar\frac{\partial}{\partial t}\Psi(t|{\bf R}) = \hat{H}_{el}(t|{\bf R})\Psi(t|{\bf R})\, ,
\end{equation}
with $\Psi({t|\bf R})$ being the time-dependent wavefunction, at nuclear coordinates ${\bf R}$.
The electronic Hamiltonian is defined as
\begin{equation}\label{H_elec}
\hat H_{e}(t|{\bf R})=\hat V_{NN}({\bf R})+\hat T_{e}+\hat V_{Ne}({\bf R})+\hat V_{ee}+ \hat V_{ext}(t),
\end{equation}
where $\hat V_{NN}$ denotes the nuclear-nuclear repulsion, $\hat T_e$ is the kinetic energy of the electrons, $\hat V_{Ne}$ denotes the attraction between nuclei and electrons, $\hat V_{ee}$ is the electron-electron repulsion, and $\hat V_{ext}$ is a time-dependent external potential, for example an oscillating laser field.

In quantum-classical dynamics, the nuclear motion evolves purely classically according to Newton's equation of motion
\begin{equation}
    F_j=m_j\ddot{{R}}_j,
\end{equation}
where $F_j$ is the force acting on nuclear degree of freedom $j$ with mass $m_j$, and $\ddot{{R}}_j$ denotes the acceleration.
In Ehrenfest dynamics, the force is evaluated as the negative derivative of an energy expectation value $\langle E \rangle$, averaged over several electronic states. In TSH, in contrast, the force is obtained as the derivative of the energy of a {\it pure} adiabatic state $m$:
\begin{equation}
    F_j=-\frac{\partial E_m(\bf{R})}{\partial R_j}.
\end{equation}
For the purpose of deriving the working equations of TSH, we expand the time-dependent electronic wavefunction in terms of the adiabatic states according to the Born-Oppenheimer expansion \cite{Domcke2004conical}
\begin{equation}\label{BO_expansion_td}
\psi(t|{\bf R})=\sum_{k} C_{k}(t) \Phi_{k}({\bf R}) \, .
\end{equation}
Here, $\Phi_{k}({\bf R})$ denote the adiabatic states. $C_k(t)$ denote the complex  time-dependent expansions coefficients that define the density matrix of the system. In TSH based on linear response TDDFT (TDDFT-SH), however, the time-dependent wavefunction (Equation \ref{BO_expansion_td}) is never directly computed; instead, only a differential equation for the state coefficients is propagated in time:
\begin{equation}\label{evolution_C}
i \dot{\mathbf{C}}=[\mathbf{H}-i \mathbf{\sigma}+\mathbf{V}_{ext}]\mathbf{C}\,,
\end{equation}
where the first-order derivative non-adiabatic coupling (NAC) \cite{Chernyak2000,Tavernelli2009,Tavernelli2009a,Domcke2004conical} is given by
\begin{equation}\label{eq:NAC}
    \sigma_{kj} = \langle \Phi_k | \frac {\partial}{\partial t} \Phi_j \rangle\ .
\end{equation}
 The Hamiltonian matrix $\mathbf{H}$ contains the adiabatic eigenenergies $H_{nm}=E_{nm}\delta_{nm}$. Within linear response TDDFT \cite{Furche2002}, the ground state energy $E_{00}$ is obtained from a ground state density functional theory (DFT) calculation, whereas for excited states $E_{nn}$ are obtained by adding the excitation energies $\omega_n$ to the ground state energy.

Typically, a photochemical reaction is initiated by absorption of light, promoting the system to an electronically excited state. Therefore, in the beginning of a TDDFT-SH simulation the system is assumed to be in a pure adiabatic state and the population of the initial state $m$ is unity ($C_m^{*}C_m=1$).
Excited state nuclear forces are then computed using linear response TDDFT \cite{Furche2002} and used to propagate the nuclear degrees of freedom on an adiabatic Born-Oppenheimer (BO) surface. At each time step of the TSH simulation, the 3$N$-dimensional Cartesian derivative coupling vector between the state of interest and neighboring electronic states is computed
\begin{equation}\label{derivative_coupling}
\mathbf{d}_{kj}(\mathbf{R})=\braket{\Phi_k(\mathbf{R})}{\nabla_{\bf R} \Phi_j(\mathbf{R})}.
\end{equation}
Using the chain rule, the NAC can be computed as the product of the nuclear velocities and the Cartesian NAC vector
\begin{equation}
    \sigma_{kj} = \dot{\textbf{R}}\cdot{\textbf{d}}_{kj}\, .
\end{equation}
Within localized basis sets the NAC vector can be efficiently calculated using the coupled perturbed Kohn-Sham equations \cite{Send2010}.
From the NAC and the state amplitudes (Equation \ref{evolution_C}), a hopping probability is computed at each time step
\begin{equation}
    g_{nm}=\frac{2\Delta t}{|C_n|^2}[\Im(C_m^*(H_{nm}-V_{ext})C_n)-\Re(c_m \sigma_{mn} C_n)]\,.
\end{equation}
The probability is compared to a random number from $\zeta \in [0,1]$ and a surface hop from state $m$ to $n$ ($m \ne n$) is accepted if and only if $\zeta$ satisfies
\begin{equation}
    \sum_{k<n} g_{mk}< \zeta < \sum_{n<k} g_{mk} .
\end{equation}
After a surface hop, the nuclear forces of the new electronic state are used to propagate the nuclear coordinates. In case the new state is the electronic ground state, the forces are computed using DFT. Furthermore, changing the electronic state requires scaling of the nuclear momenta in order to conserve the total energy of the system. According to Tully's recipe \cite{Tully1990}, scaling should be done parallel to the first order derivative NAC vector (Equation \ref{derivative_coupling}). In cases, where the NAC vector is not available, isotropic velocity scaling has been applied \cite{Doltsinis2002,Tapavicza2007}. However, proper velocity scaling has been found to be crucial for an accurate description of the dynamics in some systems \cite{Vincent2016,barbatti2021velocity}.

After having generated the initial conditions of the system of interest (see section \ref{boltzmann_ensembles}), typically several hundreds of representatives of this ensemble are used to carry out the excited state non-adiabatic simulations. 
Each trajectory will follow a distinct relaxation pathway, allowing to observe several reaction channels.

The~TDDFT-SH method has been implemented in several variations and applied to study a variety of systems \cite{Tapavicza2007,Tapavicza2011,Wiebeler2014,Wiebeler2014b}. For molecular systems, its implementation with localized basis sets \cite{Tapavicza2013,balasubramani2021TURBOMOLE} offers better efficiency compared to plane wave basis sets \cite{Tapavicza2007}.
An important advantage of localized basis sets lies in the efficient application of hybrid exchange-correlation functionals, which are crucial in the description of excited states with charge-transfer character \cite{Dreuw2003,Tapavicza2009}.

Regarding the accuracy of TDDFT, one has to be aware of 
shortcomings regarding the following issues in excited state simulations: (i) charge-transfer excitations, (ii) Rydberg excitations, (iii) conical intersections, and (iv) double excitations. While charge-transfer and Rydberg excitations can be treated sufficiently accurate with hybrid exchange-correlation functionals, it is still advisable to carefully benchmark results by comparison with more accurate wavefunction based methods, such as the CC2 method. Double excitations can be described within TDDFT using a frequency dependent exchange-correlation kernel \cite{Maitra2004}. The erroneous description of conical intersections between ground and excited states within TDDFT \cite{Levine2006,Tapavicza2008} arises from its single-reference character. It can cause problems in excited state simulations, due to the reduced dimension of the intersection space and due to discontinuous potential energy surfaces.
While practical approaches have been used to handle these problems \cite{Tapavicza2013},
recently, several TDDFT methods based on an ensemble description have been proposed to cure these deficiencies \cite{filatov2017description,lee2019conical}.

\subsection{Prediction of product quantum yields}
According to the IUPAC,
the wavelength-dependent product quantum yield (PQY) of a photochemical reaction is defined \cite{braslavsky2007glossary} as
the number of product molecules formed upon excitation at wavelength $\lambda$ ($N^{\text{P}}_{\text{molecules}}(\lambda)$) divided by the total number of photons of wavelength $\lambda$ absorbed ($N^{\text{tot}}_{\text{photons}}(\lambda)$) \cite{Thompson2018}:
\begin{equation}\label{qy}
\Phi(\lambda)=\frac{N^{\text{P}}_{\text{molecules}}(\lambda)}{N^{\text{tot}}_{\text{photons}}(\lambda)}.
\end{equation}
Hence, to predict $\Phi(\lambda)$ from first principles it is necessary to know two contributions: a) the efficiency of light absorption, which is given by the absorption cross section $\sigma(\lambda)$ or the extinction coefficient (Equation \ref{extinction}) and b) the probability that the specific product is formed once the molecule is excited. We will refer to latter probability to as the {\it branching ratio}. For conformationally flexible molecules, the macroscopic absorption cross section often strongly depends on the equilibrium distribution of conformers present in the ground state, which we have seen on the example of tachysterol (Figure \ref{fig:tachyspectra}). The branching ratio of a photochemical reaction is governed by the dynamics induced by photoexcitation, which inherently depends on the topology and shape of the excited and ground state potential energy surfaces, non-adiabatic effects, and the temperature.
Moreover, the course of a trajectory depends on the initial conformation of a molecule. Thus, one can suspect that different conformers will exhibit different photodynamics and potentially lead to different photoproducts. Consequently, in order to predict wavelength-dependent PQYs, it is necessary to account for the dependency of the molecular conformation on the absorption properties, as well as for the dependency of the molecular conformation on the photodynamical behavior.
Since TSH is based on ensemble approach that allows trajectories to follow different reaction pathways, it is well-suited to account for both contributions.

In the definition of the PQY (Equation~\ref{qy}), the number of product molecules
$N^{\text{P}}_{\text{molecules}}(\lambda)$ is equivalent to the number of photons absorbed by reactants that successfully form product $P$ ($N^{\text{P}}_{\text{photons}}(\lambda)$), assuming that that one photon is needed to produce one product molecule. This allows us to rewrite Equation~\ref{qy} as
\begin{equation}
\Phi(\lambda)=\frac{N^{\text{P}}_{\text{photons}}(\lambda)}{N^{\text{tot}}_{\text{photons}}(\lambda)}.
\end{equation}
Considering monochromatic radiation of wavelength $\lambda$,
the number of photons absorbed by a material is proportional to the oscillator strength, which is related through a line function to the absorption cross section $\sigma(\lambda)$ or, equivalently, to the extinction coefficient $\epsilon(\lambda)$ (Equation \ref{extinction}).
In the framework of TSH, $N^{\text{P}}_{\text{photons}}(\lambda)$ is proportional to the average absorption cross section of the initial structures of trajectories that successfully lead to a given product $\sigma_P(\lambda)$:
\begin{equation}\label{qy1}
        N^{\text{P}}_{\text{photons}}(\lambda)=k\frac{1}{N^{\text{tot}}_{\text{traj}}}\sum_i^{N^{\text{P}}_{\text{traj}}}\sigma_i(\lambda)=k\sigma_P(\lambda). 
\end{equation}
$N^{\text{tot}}_{\text{photons}}(\lambda)$ is proportional to the average absorption cross section of the initial structures of all trajectories $\sigma_{tot}(\lambda)$:
\begin{equation}\label{qy2}
        N^{\text{tot}}_{\text{photons}}(\lambda)=k\frac{1}{N^{\text{tot}}_{\text{traj}}}\sum_i^{N^{\text{tot}}_{\text{traj}}}\sigma_i(\lambda)=k\sigma_{tot}(\lambda).
\end{equation}

Here, $k$ is a constant relating the number of photons to the absorption cross section, and
$N^{\text{P}}_{\text{traj}}$ and $N^{\text{tot}}_{\text{traj}}$ are the number of trajectories that successfully form products and the total number of trajectories, respectively.
Hence, within TSH, the product quantum yield can be computed as
\begin{equation}\label{qy3}
        %\Phi(\lambda)=\frac{\frac{1}{N^{\text{P}}_{\text{traj}}}\sum_i^{N^{\text{P}}_{\text{traj}}}\sigma_i(\lambda)}{\frac{1}{N^{\text{tot}}_{\text{traj}}}\sum_i^{N^{\text{tot}}_{\text{traj}}}\sigma_i(\lambda)}=\frac{\sigma_P(\lambda)}{\sigma_{tot}(\lambda)}.
        \Phi(\lambda)=\frac{\sum_i^{N^{\text{P}}_{\text{traj}}}\sigma_i(\lambda)}{\sum_i^{N^{\text{tot}}_{\text{traj}}}\sigma_i(\lambda)}=\frac{\sigma_P(\lambda)}{\sigma_{tot}(\lambda)}.
\end{equation}
For the practical computation of the PQY using the outlined approach, trajectories need to be grouped according to their products formed. Then, for each group the average absorption spectrum $\sigma_P(\lambda)$ is computed by averaging the broadened spectra of each TSH initial structure. Analogously, the total spectrum $\sigma_{tot}(\lambda)$ is computed by averaging the spectra of all initial structures. Dividing the product spectrum by the total spectrum of all initial structure, according to Equation~\ref{qy3}, yields the PQY. This approach avoids binning of the initial structures according to their excitation energy.

In most TSH studies, the PQY is simply taken as the ratio of products forming trajectories \cite{Wiebeler2014,Tapavicza2011,bockmann2012enhanced,shemesh2018molecular,yue2020quantum}, which we have termed the branching ratio. In case of a wavelength independent photochemistry, the branching ratio converges to the PQY. However, in the general case the absorption energy and efficiency of the initial structures has to be taken into account to obtain PQYs according to the IUPAC definition (Equation \ref{qy}). The above outlined strategy achieves this goal.

\section{Applications}
\subsection{Photochemistry of Z-hexatriene derivatives}
\begin{figure}
    \centering
    \includegraphics{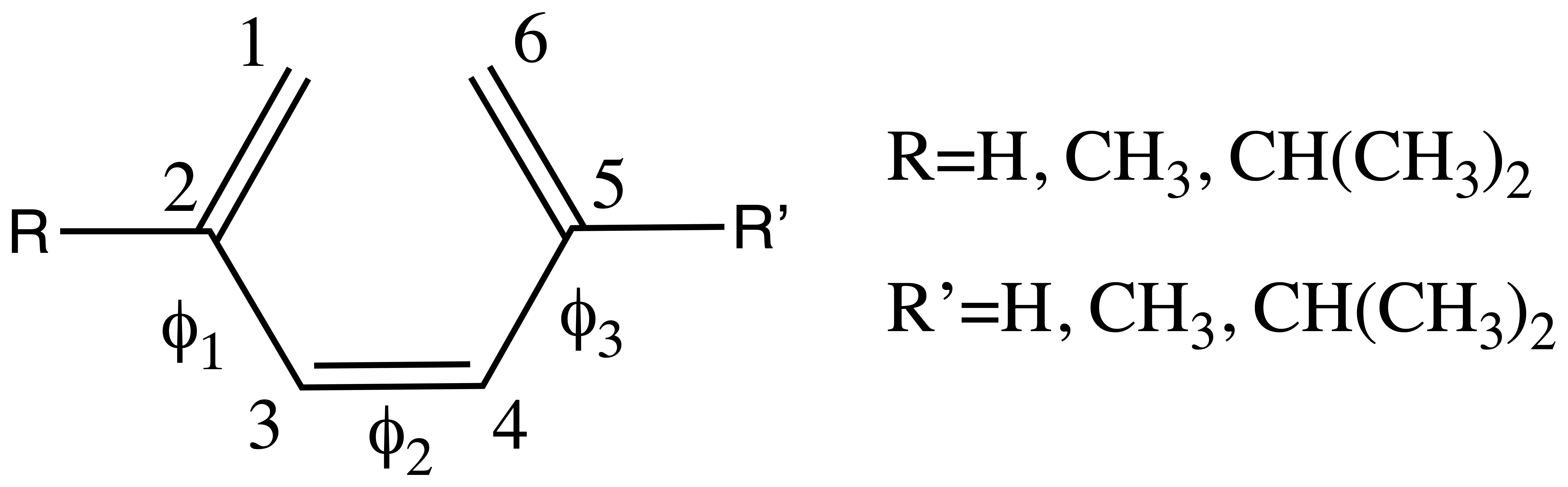}
    \caption{General structure of substituted 3Z-hexatriene derivatives. Dihedral angles $\Phi_1$, $\Phi_2$, and $\Phi_3$ are defined by carbon atoms C1-C2-C3-C4, C2-C3-C4-C5, and C3-C4-C5-C6, respectively.}
    \label{fig:definitions}
\end{figure}

\begin{figure}
    \centering
    \includegraphics[scale=0.2]{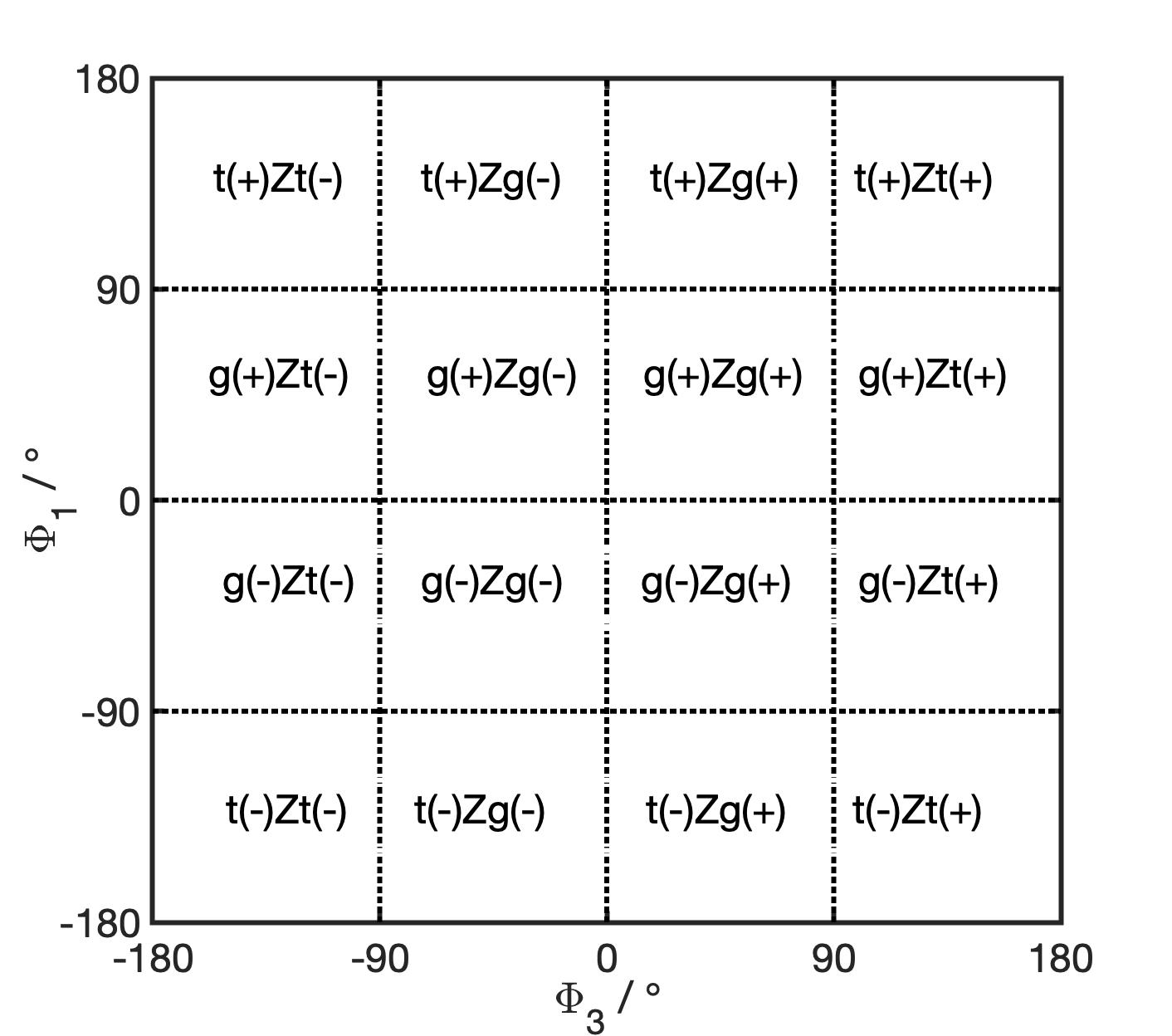}
    \caption{Nomenclature used to describe conformational isomers in 3Z-1,3,5-hexatriene derivatives. Dihedral angles $\Phi_1$ and $\Phi_3$ are defined in Figure \ref{fig:definitions}.}
    \label{fig:ramachandran}
\end{figure}

\begin{figure}
  \includegraphics[scale=0.24]{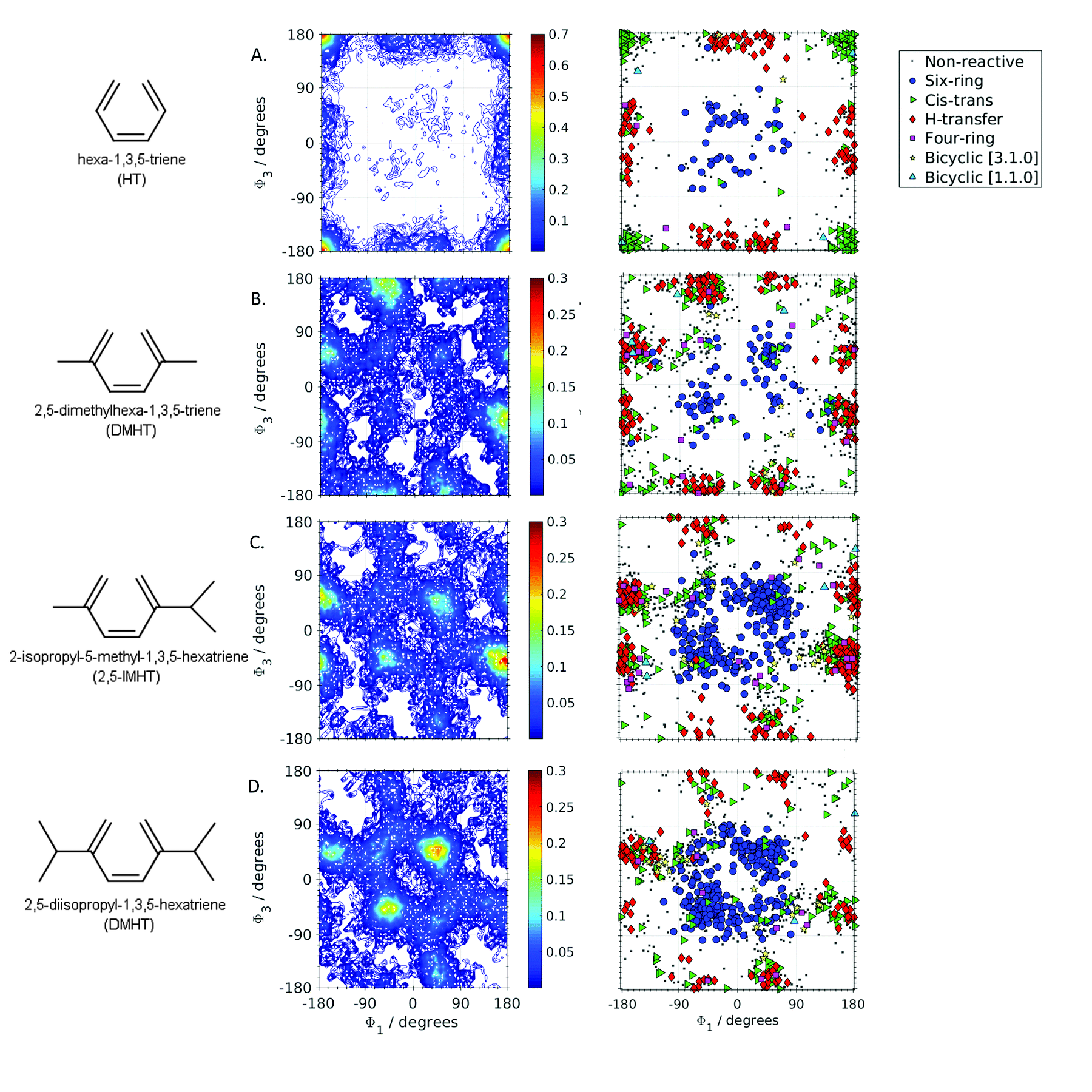}
 \caption{Relationship between ground state rotamers and photochemical product distribution: Left: Lewis drawing of hexatriene derivative. Middle: density distribution in the space of $\phi_1$ and $\phi_3$. Right: Photochemical product distribution as a function of $\phi_1$ and $\phi_3$.
Reproduced from from Tapavicza, E., Thompson, T., Redd, K. and Kim, D. (2018), Phys. Chem. Chem. Phys. 20, pp. 24807–24820 with permission from the PCCP Owner Societies.}
\label{fgr:products} 
\end{figure}

Substituted Z-hexatriene derivatives have been one of the first systems investigated that led to the hypothesis of the NEER principle \cite{Havinga1973b,Vroegop1973}.
One of the major conclusions of these studies was that the "various conformers of the ground state upon absorption of UV light yield different excited species which do not interconvert during their short (singlet excited) lifetime" \cite{Vroegop1973}. Thus each conformer is thought to produce a distinct photoproduct distribution. To investigate the validity of this hypothesis, we carried out TDDFT-SH simulations of 3Z-hexa-1,3,5-triene (HT) and several of its substituted derivatives (Figure \ref{fig:definitions}), \cite{Tapavicza2018} namely 2,5-dimethyl-HT (DMHT), 2-isopropyl-5-methyl-HT (IMHT), and 2,5,-diisopropyl-HT (DIHT), depicted in the first column of Figure \ref{fgr:products}. To generate a Boltzmann ensemble of structures for these molecules we carried out REMD based on DFT using the PBE approximation to the exchange-correlation \cite{Perdew1996}. The ensemble of structures can be presented in the space of dihedral angles $\Phi_1$ and $\Phi_3$ (see Figure \ref{fig:definitions} for definitions), similar to the well known Ramachandran plot used in biochemistry to analyze dihedral angle conformations of peptides \cite{RAMACHANDRAN196395}.
In the following, we will adopt the nomenclature of Sension et al. \cite{sofferman2021probing} for labeling the 3Z-hexatriene conformers: the structures in which both dihedral angles lie between -90 $^\circ$ and 90 $^\circ$ are labeled gZg (rather than cZc, which is also found in the literature \cite{Jacobs1979,Tapavicza2011}) to highlight their helical structure. The more planar structures, are labeled tZt, tZg, and gZt. In addition we use the sign of the dihedral angle to highlight the specific
location in the dihedral space (see Figure \ref{fig:ramachandran}), if necessary.
Analyzing the density of structures from the 300 K REMD trajectory in the dihedral space (middle column of Figure \ref{fgr:products}), we note that unsubstituted HT mainly adopts a tZt conformation, with the highest density located at the corners of the $\Phi_1/\Phi_3$-plot. To a lesser degree, also tZg and gZt conformations are present; in the  $\Phi_1/\Phi_3$-plot they are located at the borders of the plot between -90$^{\circ}$ and 90$^{\circ}$. Very few structures are present in the gZg conformation (center of the $\Phi_1/\Phi_3$-plot).
In contrast, for the substituted HT derivatives, we note that with increasing size of the substituents, more density is found in the central gZg region. The highest amount of gZg structures can be found in the case DIHT, which almost shows no tZt conformations. The steric repulsion of the bulky isopropyl substituents forces the molecules into the gZg conformation.

\begin{figure}
    \centering
    \includegraphics[scale=0.3]{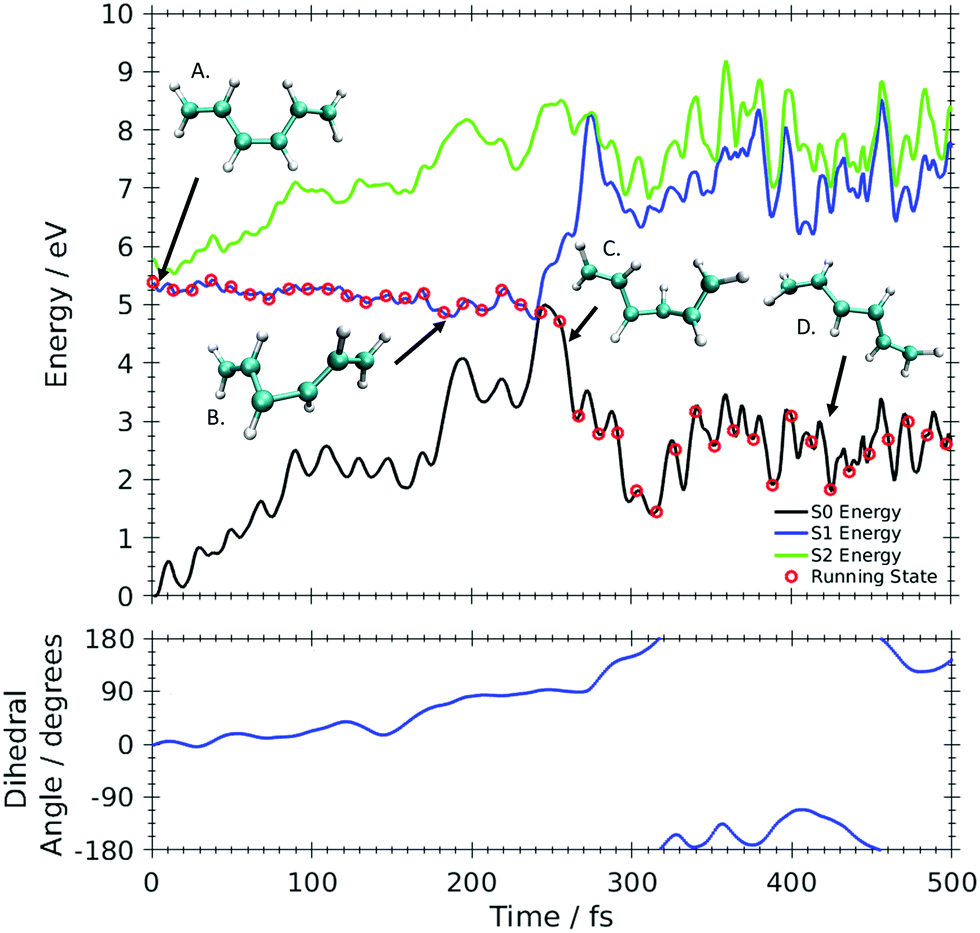}
    \caption{Upper panel: Potential energy surfaces and snapshot structures of an example trajectory that forms all-trans hexatriene via Z/E-isomerization in HT. Snapshot structures were taken at times 0 fs (A), 186 fs (B), 263 fs (C), and 429 fs (D). Lower panel: Time evolution of the dihedral angle ($\Phi_2$) of the central double bond. 
Reproduced from from Tapavicza, E., Thompson, T., Redd, K. and Kim, D. (2018), Phys. Chem. Chem. Phys. 20, pp. 24807–24820 with permission from the PCCP Owner Societies.}
    \label{fig:HT_trajectory}
\end{figure}

\begin{figure}
    \centering
    \includegraphics[scale=0.3]{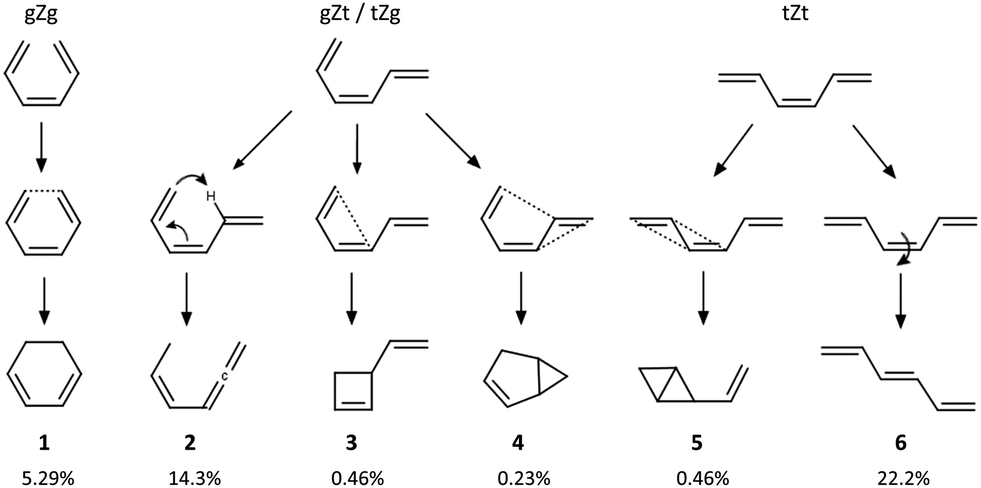}
    \caption{Reaction channels observed in TDDFT-SH simulations for unsubstituted hexatriene. 1-isopropyl-4-methyl-HT (1,4-IMHT) is not further discussed here. Branching rations for the different compounds are given in percent. 
Reproduced from from Tapavicza
, E., Thompson, T., Redd, K. and Kim, D. (2018), Phys. Chem. Chem. Phys. 20, pp. 24807–24820
 with permission from the PCCP Owner Societies.}
    \label{fig:HT_channels}
\end{figure}

\begin{figure}
    \centering
    \includegraphics[scale=0.2]{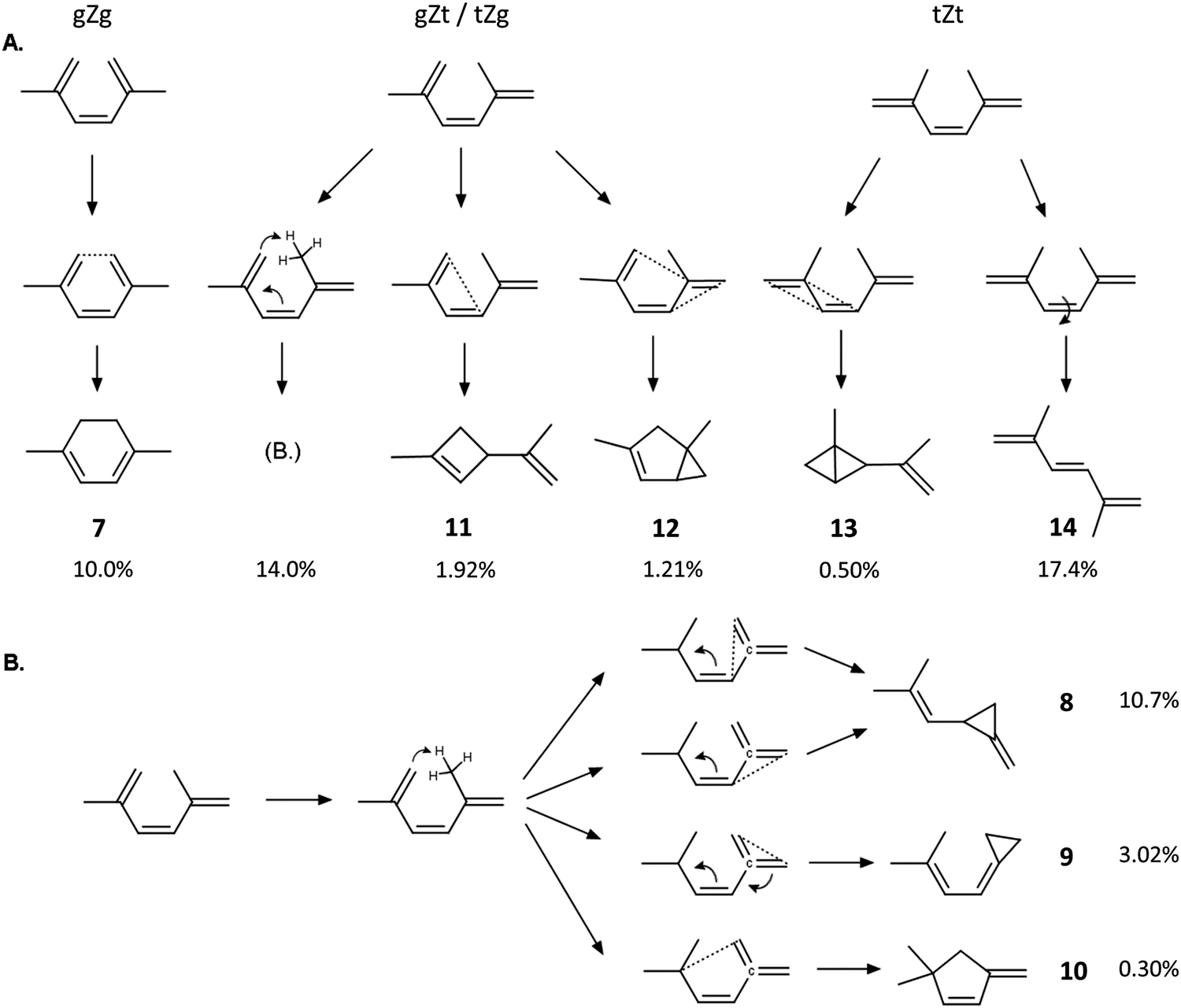}
    \caption{Reaction channels observed in TDDFT-SH simulations for 2,4-dimethyl-hexatriene. Branching rations for the different compounds are given in percent. 
Reproduced from from Tapavicza, E., Thompson, T., Redd, K. and Kim, D. (2018), Phys. Chem. Chem. Phys. 20, pp. 24807–24820 with permission from the PCCP Owner Societies.}
    \label{fig:DMHT_channels}
\end{figure}

\begin{figure}
    \centering
    \includegraphics[scale=0.3]{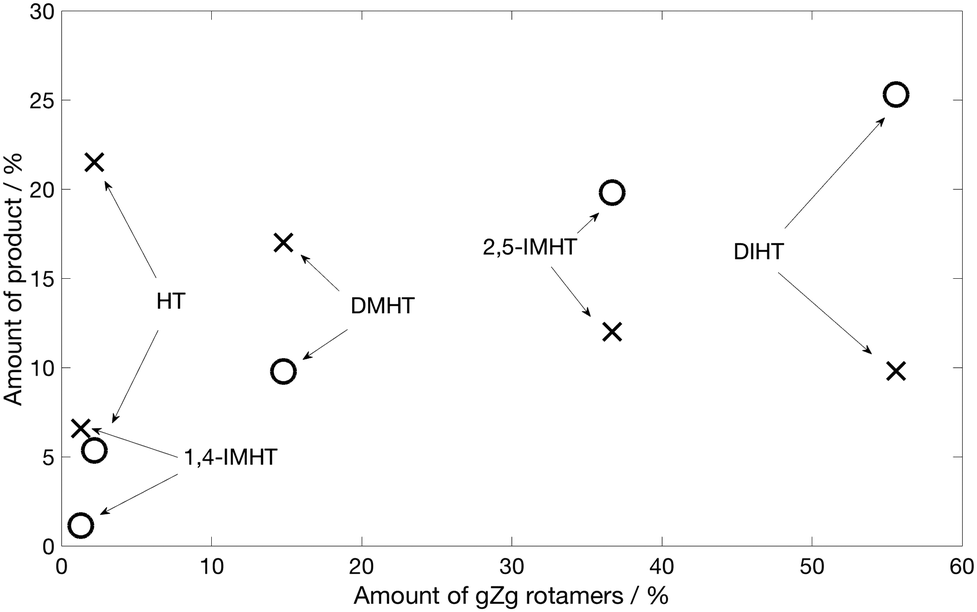}
    \caption{Branching ratio as a function of the percentage of gZg rotamers in the ground state. 
Reproduced from from Tapavicza, E., Thompson, T., Redd, K. and Kim, D. (2018), Phys. Chem. Chem. Phys. 20, pp. 24807–24820 with permission from the PCCP Owner Societies.}
    \label{fig:QYvsconformation}
\end{figure}

To investigate the dependency of the photochemical reactivity on the dihedral angle conformation, we carried out TDDFT-SH simulations with each of the HT derivatives.
Between 800 - 1400 starting structures and velocities were chosen randomly from the ensemble of structures generated by REMD (middle column of Figure \ref{fgr:products}).
 In TDDFT-SH simulations, we used the hybrid PBE0 approximation to the exchange-correlation functional \cite{Perdew1996a}, which contains one fourth exact (Hartree-Fock like) exchange. To obtain accurate excitation energies and potential energies, usage of PBE0 in TDDFT (TDBPE0) is crucial.
At time zero of the TDDFT-SH simulations, the molecules were prepared in the first excited singlet state (S$_1$), which possesses the highest oscillator strength among the ten lowest excitations within TDPBE0.
The nuclear coordinates are then propagated using the excited state nuclear forces of this state. An example trajectory of HT, undergoing a Z/E-isomerization of the central double bond, is shown in Figure \ref{fig:HT_trajectory}. This is the main reaction channel of HT, with a branching ratio of 22 \% (Figure \ref{fig:HT_channels}). Besides minor reaction channels forming products {\bf 3}--{\bf 5}, we also find a photoinduced [1,5]-hydrogen shift forming product {\bf 2} and an electrocyclic ring-opening forming cyclohexadiene (CHD), product {\bf 1}. Analyzing the trajectories of the substituted DMHT (Figure \ref{fig:DMHT_channels}), we find an increased amount of electrocyclic ring-closing reaction relative to unsubstituted HT (formation of product {\bf 7}). In contrast, the amount of Z/E-isomerization is reduced compared to unsubstituted HT. For the other substituted derivatives IMHT and DIHT, the trend continues exhibiting increased amount ring-closing reactions and decreased Z/E-isomerizations (Figure \ref{fig:QYvsconformation}).

\begin{figure}
    \centering
    \includegraphics[scale=0.3]{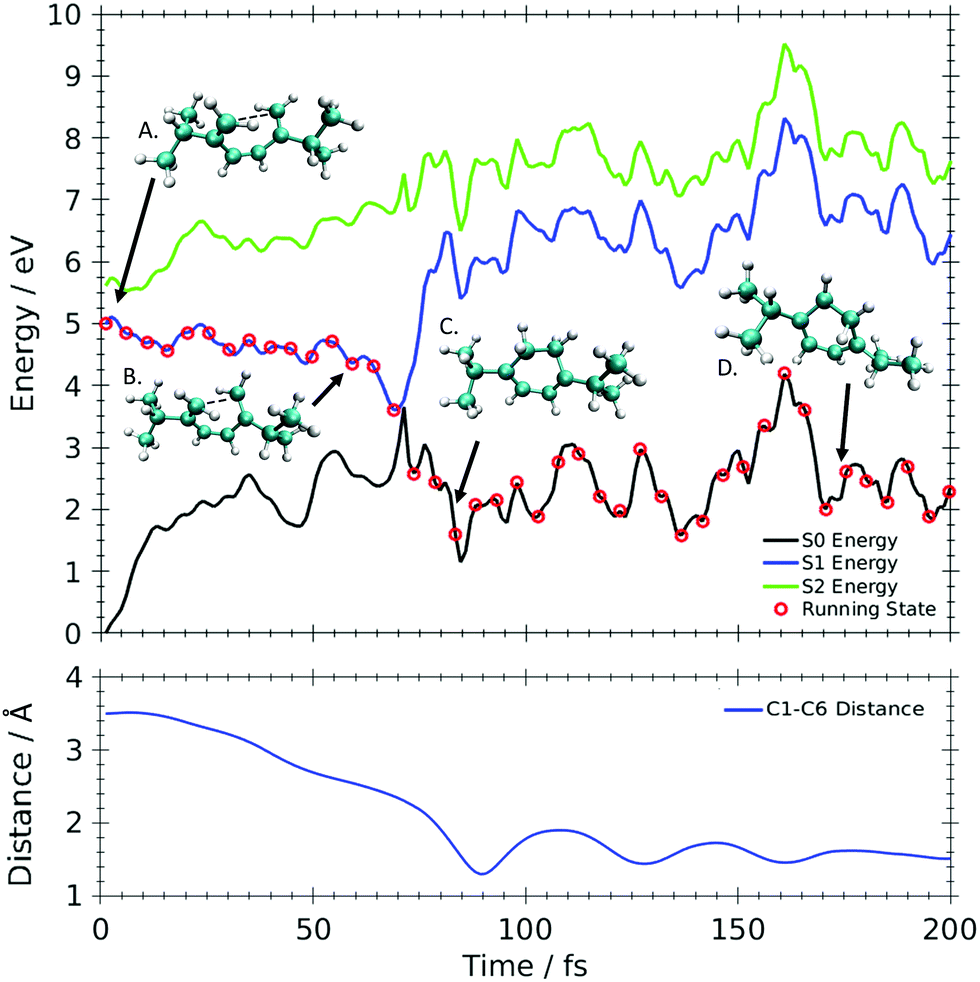}
    \caption{Upper panel: Potential energy surfaces and snapshot structures of an example trajectory of DIHT forming 1,4-diisopropyl-cyclohexadiene. Total simulation times of the snapshot structures are 0 fs (A), 60 fs (B), 83 fs (C), and 177 fs (D). Lower panel: C1–C6 distance as a function of time. Atom numbers are defined in Figure \ref{fig:definitions}. 
Reproduced from from Tapavicza, E., Thompson, T., Redd, K. and Kim, D. (2018), Phys. Chem. Chem. Phys. 20, pp. 24807–24820 with permission from the PCCP Owner Societies.}
    \label{CHD_formation}
\end{figure}

\begin{figure}
    \centering
    \includegraphics[scale=0.3]{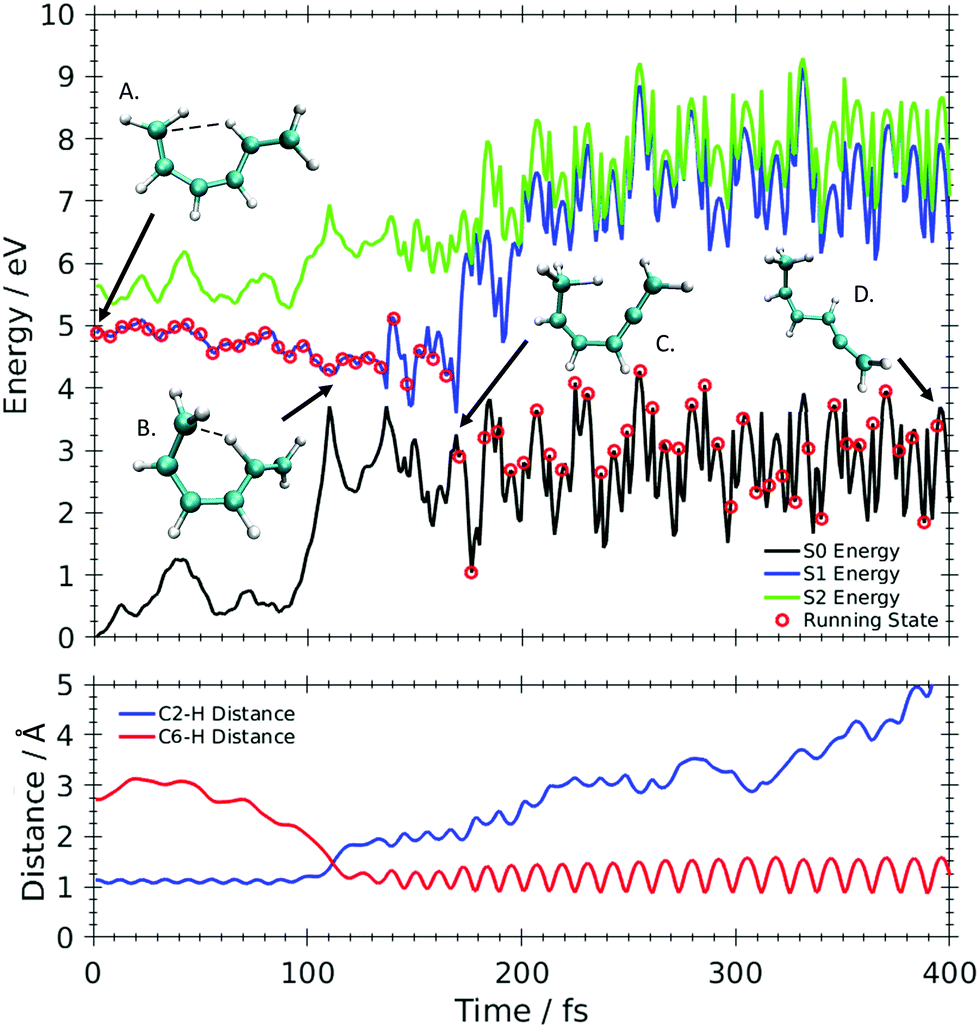}
    \caption{Upper panel: Potential energy surfaces and snapshot structures of an example trajectory that forms allene {\bf 2} via [1,5]-sigmatropic hydrogen shift in HT. Lower panel: Time evolution of the distance between carbon C2 and hydrogen (blue) and carbon C6 and hydrogen (red). 
Reproduced from from Tapavicza, E., Thompson, T., Redd, K. and Kim, D. (2018), Phys. Chem. Chem. Phys. 20, pp. 24807–24820 with permission from the PCCP Owner Societies.}
    \label{H_trans_HT}
\end{figure}

To understand the relationship between initial conformation and the outcome of the simulation, we plotted the reaction products as a function the initial dihedral angle conformation (right column in Figure \ref{fgr:products}). This analysis shows that CHD derivatives formed by electrocylic ring-closing only originate from initial structures in the gZg conformation.  This behavior can be explained by the fact that only in the gZg conformations the critical carbon atoms (C1 and C6, according to Figure \ref{fig:definitions}) that form the new $\sigma$-bond are close enough for this reaction to happen. Because of the short excited state lifetime, ranging from 115-150 fs, no major changes in dihedral angle conformation occur, which is confirmed by the example trajectory of DIHT in Figure \ref{CHD_formation}.
From Figure \ref{fgr:products}, we also infer that hydrogen-transfer is a typical reaction for tZg and gZt isomers.   This behavior is also easily understood from structural considerations: only in these two conformations the critical hydrogen atom is located in close vicinity of one of the unsaturated carbon atoms C1 or C6, as can be seen from the second reaction channel from the left in Figure \ref{fig:HT_channels} and from channel B in Figure \ref{fig:DMHT_channels}. Inspecting a trajectory of photoexcited tZg HT confirms this behavior (Figure \ref{H_trans_HT}). Also here, the initial conformation is largely conserved during the excited state dynamics.
As another major trend we observe that Z/E-isomerizations occurs mainly from tZt conformers. This is not directly understandable from structural consideration. While the isomerization of the central double bond could occur more easily in the tZt conformer, it is also possible that in the tZt conformation there is no competition with the other two mentioned reaction channels (namely electrocylic ring-closing and sigmatropic hydrogen shift) and consequently the probability of Z/E-isomerization increases. Comparing all derivatives, we can clearly see a positive, almost linear correlation between the amount of gZg conformers in the ground state and the amount CHD derivatives formed (Figure \ref{fig:QYvsconformation}). In contrast, the amount of Z/E-isomerization decreases linearly between the amount of gZg conformers in the ground state. In summary, we can say that in the case of the substituted Z-hexatriene derivatives, simulation show that it is indeed the composition of ground state rotamers at the time of photoexcitation that determines the outcome of the photochemical reaction, confirming  the validity of the NEER principle. No major rearrangements of the dihedral angle conformation is observed during the short excited state lifetime of the presented molecules.

\begin{figure}
    \centering
    \includegraphics[scale=0.4]{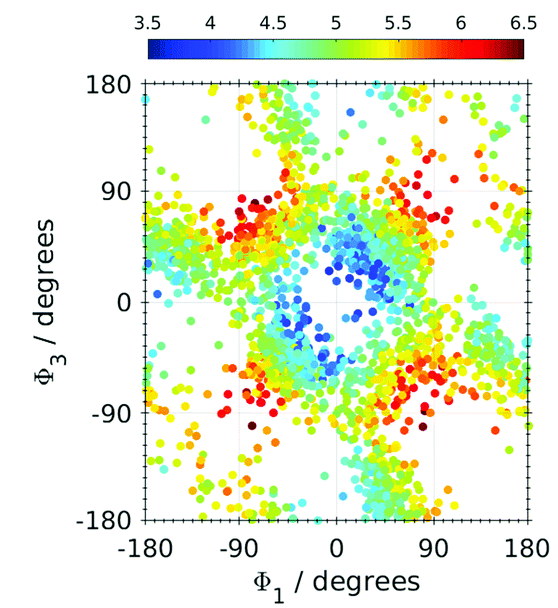}
    \caption{ADC(2)/COSMO S$_1$ $\leftarrow$ S$_0$ excitation energies of DIHT in eV as a function of dihedral angles $\Phi_1$ and $\Phi_3$. Reproduced (adapted) from Tapavicza, E., Thompson, T., Redd, K. and Kim, D. (2018), Phys. Chem. Chem. Phys. 20, pp. 24807–24820 with permission from the PCCP Owner Societies.}
    \label{fig:DIHT_exci}
\end{figure}

Another aspect that follows from this principle is the characteristic wavelength dependency caused by the fact that the different ground state rotamers exhibit distinct absorption properties.
Inspecting the dihedral angle dependency of the S$_1$ excitation energies of DIHT (Figure \ref{fig:DIHT_exci}), we see that the gZg conformers in the center of the plot exhibit lower excitation energies compared to the remaining ensemble of conformers; all substituted HT derivatives discussed here follow this trend.
As discussed above, these are the structures in which the distance between carbon C1 and C6 is close enough to form a covalent bond after photoexcitation. Excitation at low energies selectively excites those gZg conformers and, consequently, excitation at low energies increases the probability of inducing ring-closing reactions. In turn, selective excitation of tZt structures, which can be achieved by excitation at higher energies, will lead to an increase in probability of inducing the Z/E-isomerization. Hence, excitation at higher energies leads predominantly to Z/E-isomerization. This pattern is characteristic for all 3Z-hexatriene derivatives studied here, and, as we will see in the next section, it also applies to the 3Z-hexatriene derivative previtamin D.

\subsection{Vitamin D photochemistry}\label{vitamin_D}
\begin{figure}
    \centering
    \includegraphics[scale=0.2]{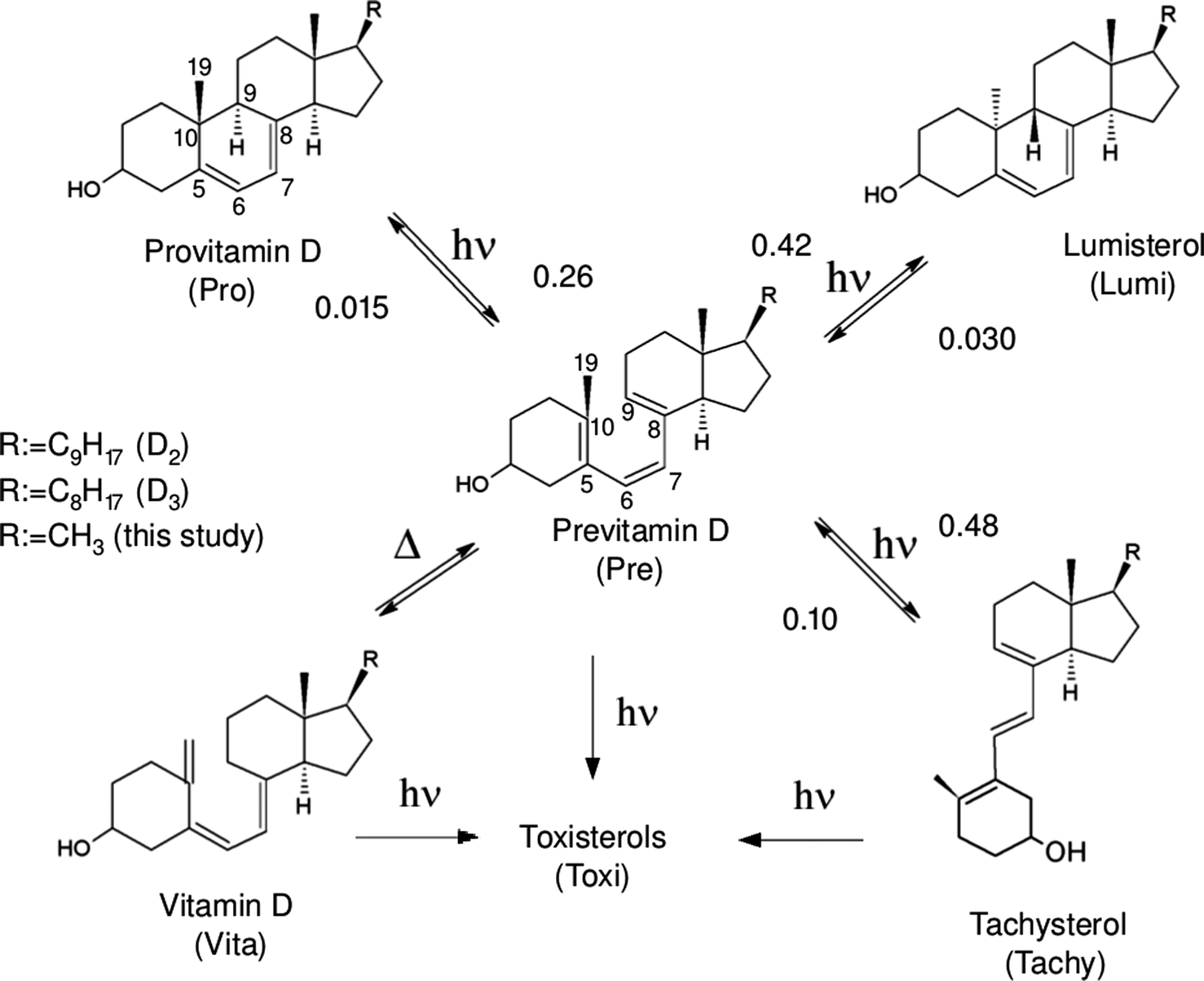}
    \caption{Photochemical and thermal reactions involved in the natural photosynthesis of vitamin D. Quantum yields measured at 253 nm are given on the arrows \protect\cite{Havinga1973}. 
Reproduced from Cisneros, C., Thompson, T., Baluyot, N., Smith, A. C. and Tapavicza, E. (2017), Phys. Chem. Chem. Phys. 19, pp. 5763–5777 with permission from the PCCP Owner Societies.}
    \label{fig:hub}
\end{figure}

\begin{figure}
    \centering
    \includegraphics[scale=0.5]{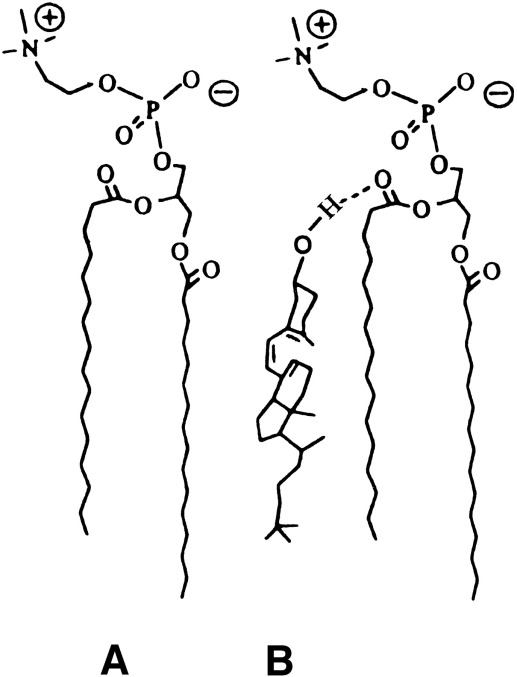}
    \caption{Proposed structural model for the localization of the gZg-previtamin D3 in a phospholipids membrane. 
   Picture taken from Tian, X. Q. and Holick, M. F. (1999), J. Biol. Chem. 274, 7, 4174-4179, licensed under Creative Commons CC-BY.}
    \label{fig:Pre_DPPC}
\end{figure}

\begin{figure}
    \centering
    \includegraphics[scale=0.2]{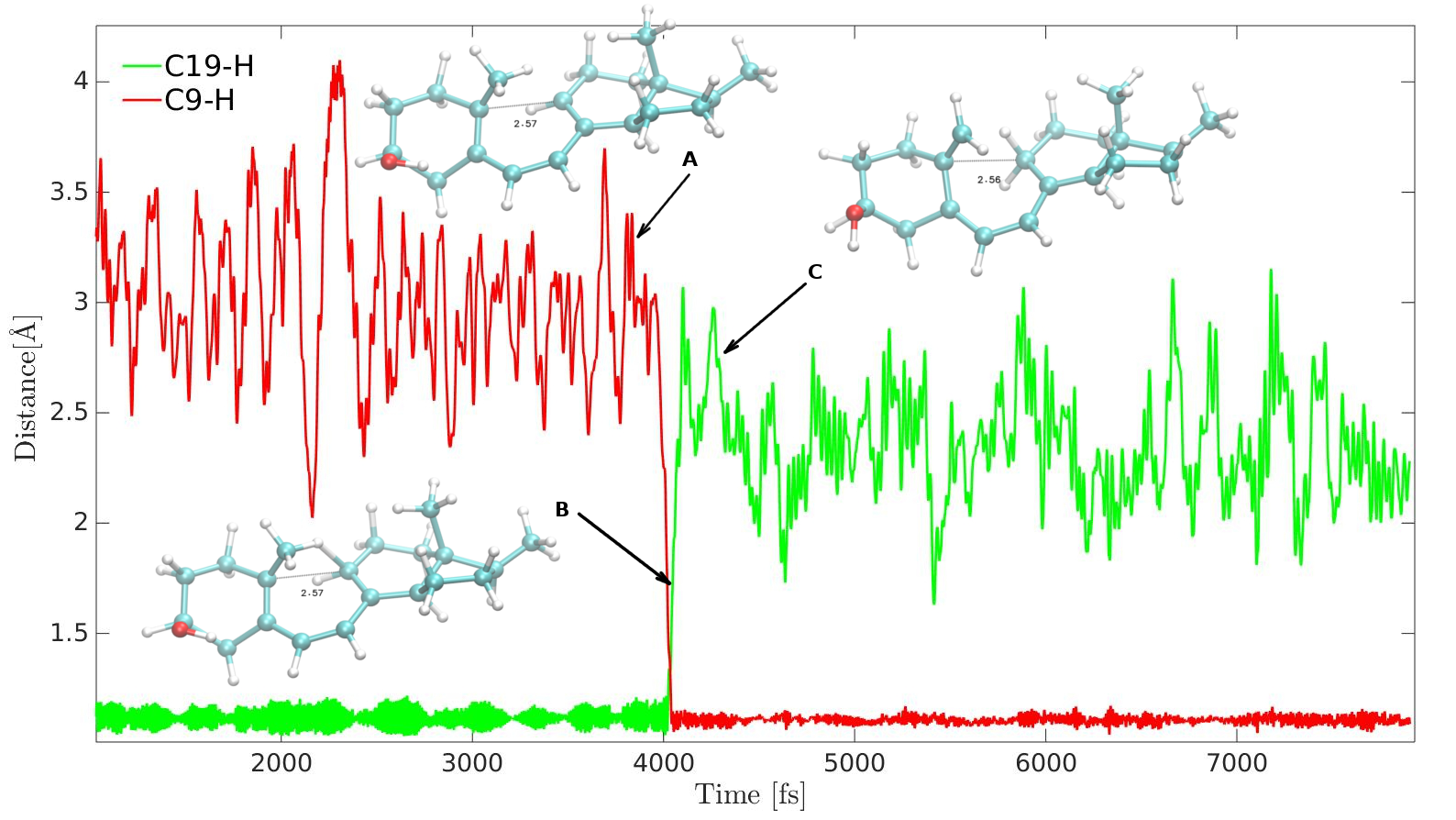}
    \caption{Results of a ground state AIMD simulation of the hydrogen transfer dynamics of the reaction previtamin D $\rightarrow$ vitamin D. Red: distance between C19 and the critical hydrogen atom, green: distance between C9 and the critical hydrogen. Atom numbers as defined in Figure \ref{fig:hub}. Molecule A represents Pre before hydrogen transfer, Structure B is the transition state, and Structure C is vitamin D after hydrogen transfer. The simulation has been carried out and a temperature of 800 K. The C9-C19 distance has been constrained to (2.56 $\pm$ 0.2) \AA\, to keep the molecule in the gZg conformation, mimicking the effect of the biological membrane.}
    \label{Htrans_Pre}
\end{figure}

Vitamin D is an important prohormone that is involved in a variety of biological processes, including calcium absorption from the blood stream \cite{Holick2003}, prevention of cancer \cite{Garland2009} and autoimmune diseases \cite{Wang2009}. It has also been found to have protective effects against acute respiratory infections \cite{martineau2020vitamin}. The largest part of the world population obtain their major part of vitamin D through natural photoproduction in the epidermal skin cells, induced by ultraviolet (UVB) light.
Natural vitamin D photosynthesis is a prominent paradigm for a conformationally controlled photochemical system in biology. The initial step of vitamin D photosynthesis is the photo-induced 6$\pi$-electrocyclic ring-opening of the central hexadiene unit in provitamin D (Pro), forming previtamin D (Pre), as indicated in the Figure \ref{fig:hub}. A biexponential excited state decay has been determined with lifetimes ranging from a few hundreds of femtoseconds to a few picoseconds \cite{Fuss1996,Anderson1999,Tang2011,MeyerIlse2012}. According to the Woodward-Hoffmann rules \cite{Woodward1965}, the ring-opening occurs in a conrotatory fashion. Its quantum yield has been determined to be between 20-65~\% \cite{Norval2010}. While the optimal wavelength for Pro/Pre conversion is around 290 nm, no pronounced wavelength dependency has been observed for this reaction \cite{vanDijk2016}.
After successful ring-opening, Pre is formed in the gZg conformation, which is thought to rapidly undergo rotational isomerization to tZg conformers \cite{sofferman2021probing}. Eventually, an equilibrium distribution of six different major conformers is formed \cite{Dmitrenko1999,Tapavicza2011}, each of which can further be divided into structures having the OH group in axial or equatorial position \cite{bayda2017lumisterol}. The time until an equilibrium distribution of conformers is adopted is about 100 ps and depends on the solvent viscosity \cite{Tang2011}.
The formed Pre can react thermally to vitamin D via a [1,7]-sigmatropic hydrogen shift (Figure \ref{fig:hub}). However, this process only occurs if carbon C9 and C19 (see Figure \ref{fig:hub} for atom numbers) are in close vicinity to each other, which is only possible in the gZg conformation. Furthermore, Pre can also undergo several other photochemical reactions, which are highly wavelength dependent and conformationally controlled \cite{Tapavicza2011}: excitation at the red side of the spectrum predominantly leads to 6$\pi$-electrocyclic ring-closing, forming either Pro or lumisterol (Lumi), a stereoisomer of Pro (Figure \ref{fig:hub}). Prolonged irradiation and excitation at the blue side of the spectrum predominantly leads to tachysterol (Tachy) formation. Besides these major photoproducts, a variety of so-called {\it toxisterols} have been determined \cite{Boomsma1975,Boomsma1977overirradiation,Jacobs1979}. Their biological role and further metabolism is largely unknown.
In living organism, these photochemical and thermal processes occur in the phospholipid bilayer of epidermal skin cells. It has been found, that both, the presence of the membrane and the spectral composition of the irradiation source strongly affects quantum yields of the involved reactions \cite{MacLaughlin1982}.
For instance, Pre formation continuously increases with increasing radiation density if monochromatic light at 295 nm is used. In contrast, if a solar simulator is used, Pre formation only increases at low radiation density; at higher radiation density Pre formation plateaus \cite{MacLaughlin1982}. Similar variations have also been found for Lumi and Tachy formation. Obviously, an intrinsic regulation occurs due to the spectral composition of the sun, preventing overproduction of vitamin D upon prolonged sun exposure. The same study \cite{MacLaughlin1982} also found slight decrease in photochemical Pre formation in skin cells compared to tetrahydrofuran solution. In contrast, the reaction rate of the isomerization of Pre to vitamin D has been found to be enhanced by a factor of 10 in epidermal skin cells and in artificial dipalmitoylphosphatidylcholine (DPPC) liposomes compared to in organic solutions \cite{holick1995evolutionary,Tian1999}.
It has been hypothesized that this increase in vitamin D formation is due to the "cholesterol-like" interactions between Pre and the phospholipid molecules that
trap the molecule in the gZg conformation (Figure \ref{fig:Pre_DPPC}), facilitating the [1,7]-sigmatropic hydrogen transfer reaction.
According to theoretical studies \cite{meana2012tunneling}, both g(+)Zg(+) and g(-)Zg(-) can convert to vitamin D.
In constrained ground state AIMD simulations we could confirm this reaction to happen in the g(-)Zg(-) conformation at a distance of about 2.56 {\AA } between carbons C9 and C19 (Figure \ref{Htrans_Pre}).

\begin{figure}
    \centering
    \includegraphics[scale=0.5]{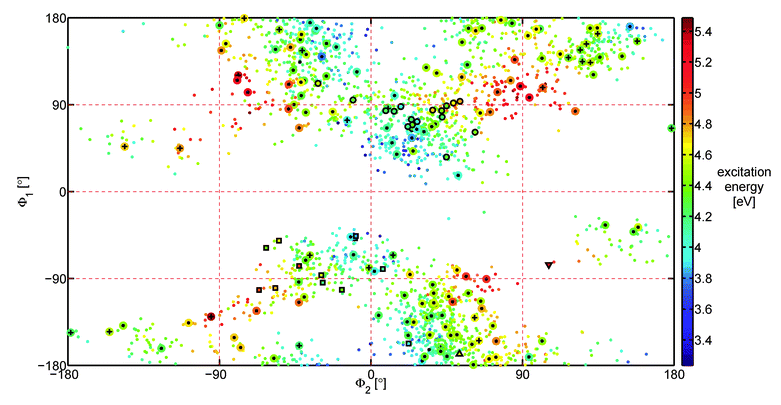}
    \caption{Excitation energy and photoproduct as a function of torsional angles $\Phi_1$ and $\Phi_2$ of Pre. The color code indicates the S$_1$ excitation energy in eV. Unreactive trajectories are indicated by a black dot. Trajectories that led to ring-closure are indicated by squares (provitamin D formation) or by circles (lumisterol formation). Crosses denote tachysterol formation via Z/E isomerization. $\Delta$ indicates hydrogen shift, forming vitamin D and $\nabla$ indicates toxisterol formation. 
Reproduced from Tapavicza, E., Meyer, A. M. and Furche, F. (2011), Phys. Chem. Chem. Phys. 13, 20986-20998 with permission from PCCP Owner Societies.  }
    \label{fig:Pre_phi_exci}
\end{figure}

Returning to the purely photochemical reactions in the vitamin D system, the secosteroids Pre and Tachy exhibit an interesting wavelength dependent photochemistry due to their flexible open-ring structure.
REMD simulations of Pre exhibit a particular dihedral angle conformation with six high density spots \cite{Tapavicza2011} corresponding to the six mentioned minimum energy structures \cite{Dmitrenko2001,bayda2017lumisterol}.
Similar to the example of tachysterol (Figure \ref{fig:tachyosc}), also Pre exhibits a conformational dependency of the absorption, which is obtained by computing excitation energies for snapshot structures from REMD simulations (Figure \ref{fig:Pre_phi_exci}).
Analogously to the Z-hexatriene derivatives, the wavelength-dependent photochemistry can be probed by correlating the S$_1$ excitation energy of a snapshot structure from the REMD ensemble with its photoproduct formed in TDDFT-SH simulations. As we can see from Figure \ref{fig:Pre_phi_exci}, the ring-closing reaction forming Lumi occurs mainly for g(+)Zg(+) conformers,
whereas the ring-closing with opposite helicality (forming Pro), occurs mainly for g(-)Zg(-) conformers.
The fact that gZg conformers on average exhibit lower S$_1$ excitation energies explains why ring-closing reaction occur mainly upon excitation at the red side of Pre's spectrum.
In contrast, Tachy formation via Z/E-isomerization is prone to occur in t(+)Zt(+) conformers, which exhibit on average higher excitation energies. With this behavior, Pre follows the same trend as the Z-hexatriene derivatives.

\begin{figure}
    \centering
    \includegraphics[scale=0.25]{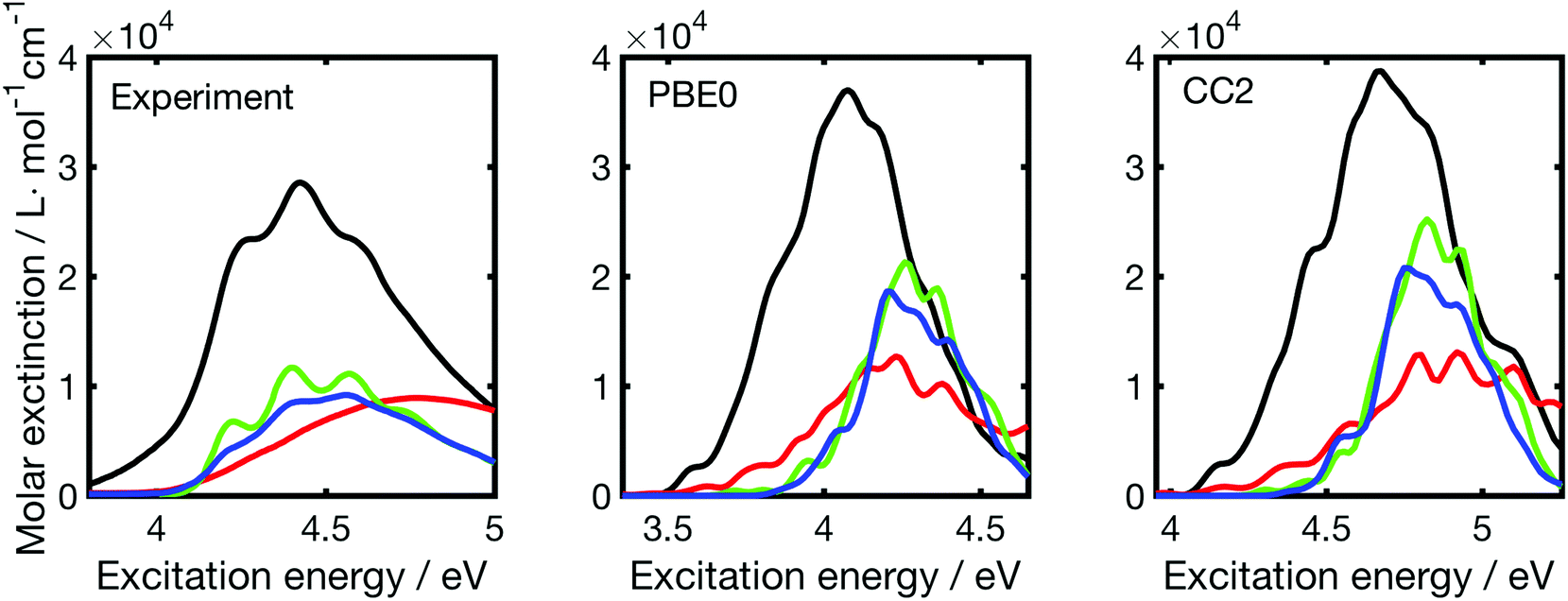}
    \caption{Comparison between the experimental absorption spectra measured in ether\protect\cite{MacLaughlin1982} (left) and the spectra calculated by TDPBE0 (middle) and CC2 (right). Tachysterol: black, provitamin D: green, previtamin D: red, lumisterol: blue. Reproduced from Cisneros, C., Thompson, T., Baluyot, N., Smith, A. C. and Tapavicza, E. (2017), Phys. Chem. Chem. Phys. 19, pp. 5763–5777 with permission from the PCCP Owner Societies.}
    \label{fig:DPI_spectra}
\end{figure}

While the photochemistry of Pro and Pre have been investigated extensively, the photochemistry and biological role of the E-hexatriene Tachy has attracted less attention. However, biological studies have shown that a variety of Tachy derivatives exhibit binding to the human vitamin D receptor \cite{suda197025,maestro2019vitamin}.
Furthermore, recently it has been hypothesized that Tachy potentially constitutes a vitamin D reservoir that can be tapped at long wavelength UV radiation and could constitute an important route to produce vitamin D in winter months, where short wavelength UV light is limited \cite{andreo2015generation}.
The possibility of inducing photochemical reactions at longer wavelengths in Tachy is already suggested by its absorption spectrum, which extends far more to the red than the spectra of the vitamin D derivatives Pro, Pre, and Lumi (Figure \ref{fig:DPI_spectra}).
 As we can see from Figure \ref{fig:tachyspectra}, the overlap with the solar radiation (brown) is largest with the spectrum of the tEt rotamers; this has a specific impact on photoreactivity of tachysterol.
Compared to the other vitamin D derivatives, Tachy's absorption spectrum exhibits also larger extinction coefficients, covering the entire absorption spectra of the other compounds. This behavior is properly described by both theoretical methods, TDDFT and CC2. However, the absorption spectrum alone does not allow to assess its photochemical reactivity.

\begin{figure}
    \centering
    \includegraphics[scale=0.25]{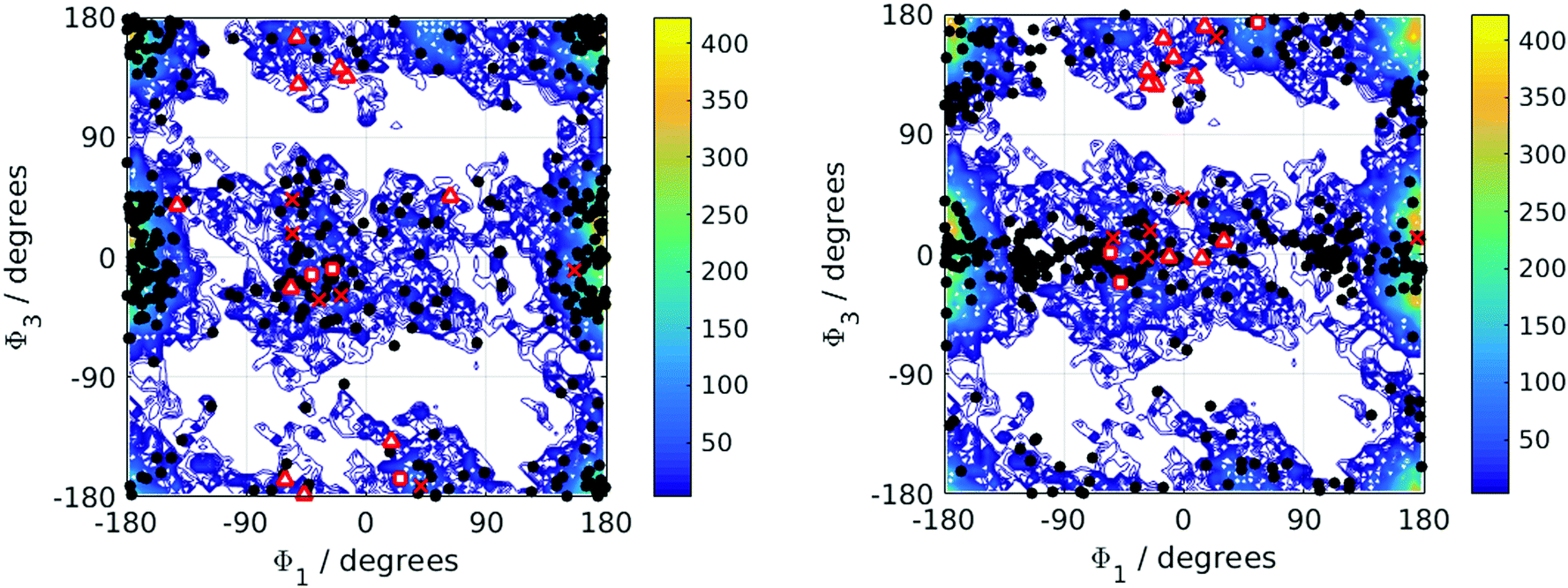}
    \caption{Distribution of rotamers as a function of the dihedral angles $\Phi_1$ and $\Phi_3$ in the ground state equilibrium of tachysterol obtained from REMD at 300 K is indicated by the contours left and right. Number of structures per 10$^{\circ}$ $\times$ 10$^{\circ}$ are indicated by the color code. Left: Position of TDDFT-SH starting structures are indicated by the symbols. Right: Position of TDDFT-SH structures at the time of the surface hop are indicated by the symbols. Black point: unreactive (product Tachy); cross: trans–cis isomerization (product Pre), triangle: hydrogen transfer (toxisterol D1); square: thermal cyclobutene (CB) formation (product: CB-toxisterol). 
Reproduced from Cisneros, C., Thompson, T., Baluyot, N., Smith, A. C. and Tapavicza, E. (2017), Phys. Chem. Chem. Phys. 19, pp. 5763–5777 with permission from the PCCP Owner Societies.}
    \label{fig:tachy_REMD}
\end{figure}

\begin{figure}
    \centering
    \includegraphics[scale=0.24]{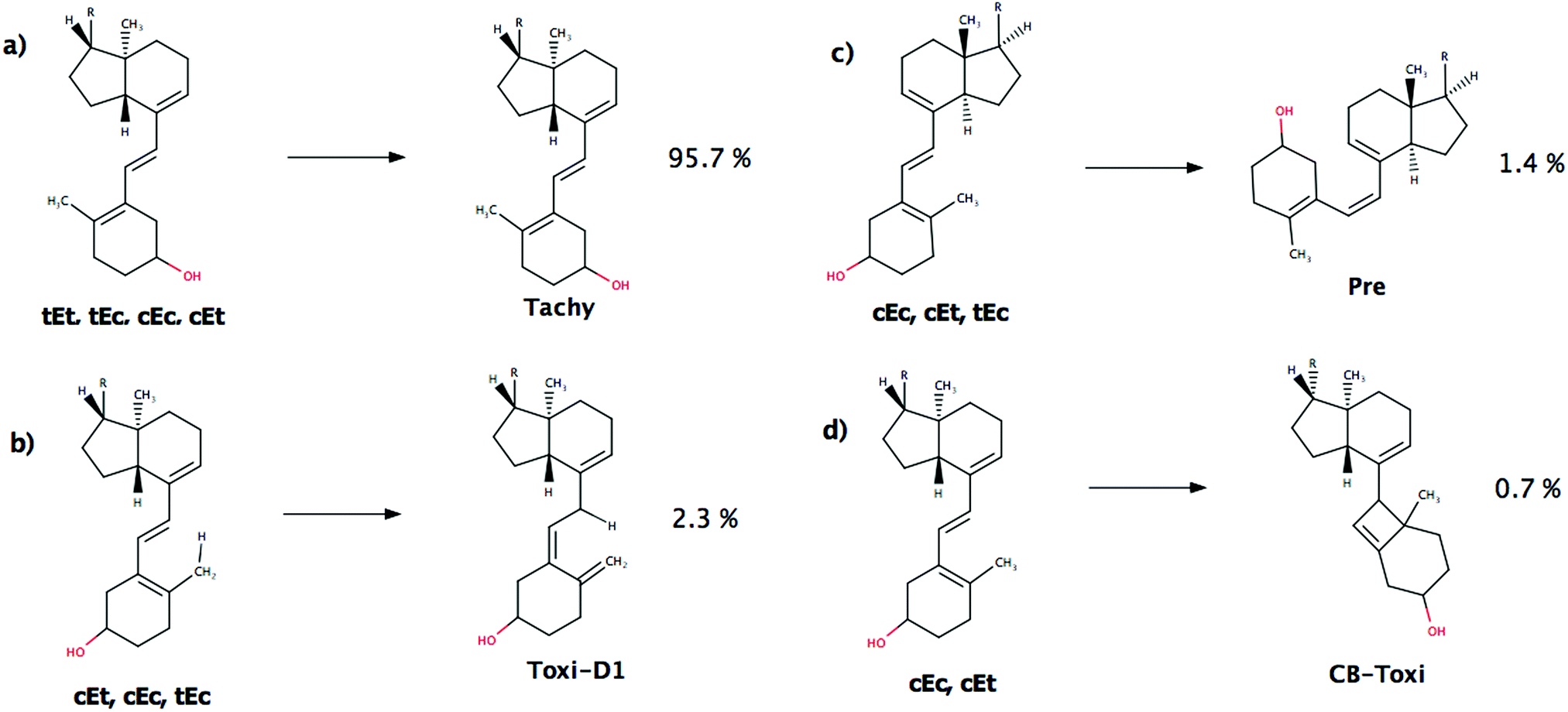}
    \caption{Overview of the observed photoreactions of tachysterol: (a) unreactive, (b) [1,5]-sigmatropic hydrogen transfer (toxisterol D1 formation), (c) hula-twist double bond isomerization (previtamin D formation), (d) thermal 2 + 2 electrocyclization (toxisterol-CB formation). Only the main rotamer in which the reaction has been observed is sketched, other rotamers in which the reaction was observed are listed below the reactant. The branching ratios for each reaction channel are given. 
Reproduced from Cisneros, C., Thompson, T., Baluyot, N., Smith, A. C. and Tapavicza, E. (2017), Phys. Chem. Chem. Phys. 19, pp. 5763–5777 with permission from the PCCP Owner Societies.}
    \label{fig:tach_trajectories}
\end{figure}

To investigate the photochemical behavior of Tachy, we carried out TDDFT-SH simulations using initial structures from an ensemble of structures generated by REMD (Figure \ref{fig:tachy_REMD}).\cite{Cisneros2017} Simulations show, that to a large part (95.7 \%) trajectories decay to the ground state without chemical transformation (Figure \ref{fig:tach_trajectories}).
Interestingly, all trajectories starting from tEt structures decay unreactively and only tEc, cEc, tEc undergo photochemical transformations to a small degree (Figure \ref{fig:tach_trajectories}).
Pre formation, which is crucial for vitamin D formation, mainly originates from cEc conformers upon excitation at the red side of the spectrum. This finding confirms the possibility that vitamin D could be formed in winter where short UV light is limited. Furthermore, we find a cyclobutene (CB) derivative formed from cEt and cEc initial structures at the red side of the spectrum. CB toxisterols have been found to thermalize to Pre \cite{Boomsma1977overirradiation} and therefore they also possibly constitute a source of vitamin D in winter.
Furthermore, we also find [1,5]-sigmatropic hydrogen transfer in a few trajectories, which for steric reasons only occurs in the cEt or cEc conformation.

Comparing the dihedral angle conformation of the initial structures of the trajectories with their conformation at the time of the surface hop, we see that the distribution of conformers has changed substantially during the excited state dynamics. Interestingly, we observe one trajectory undergoing [1,5]-sigmatropic hydrogen transfer, in which the initial tEc conformation first transformed to the cEc conformation. For steric reasons, this hydrogen transfer would not have been possible in the tEc conformation. This behavior violates the NEER principle. The anti-NEER behavior can be explained with Tachy's relatively long excited state lifetime (882 fs) compared
to the Z-hexatriene derivatives (115-150 fs).
The long lifetime allows the molecule to adopt an equilibrium of conformers corresponding to the excited state potential energy surface.

\subsection{Wavelength-dependent product quantum yields in Z-hexatriene derivatives}
\begin{figure}
    \centering
    \includegraphics[scale=1.8]{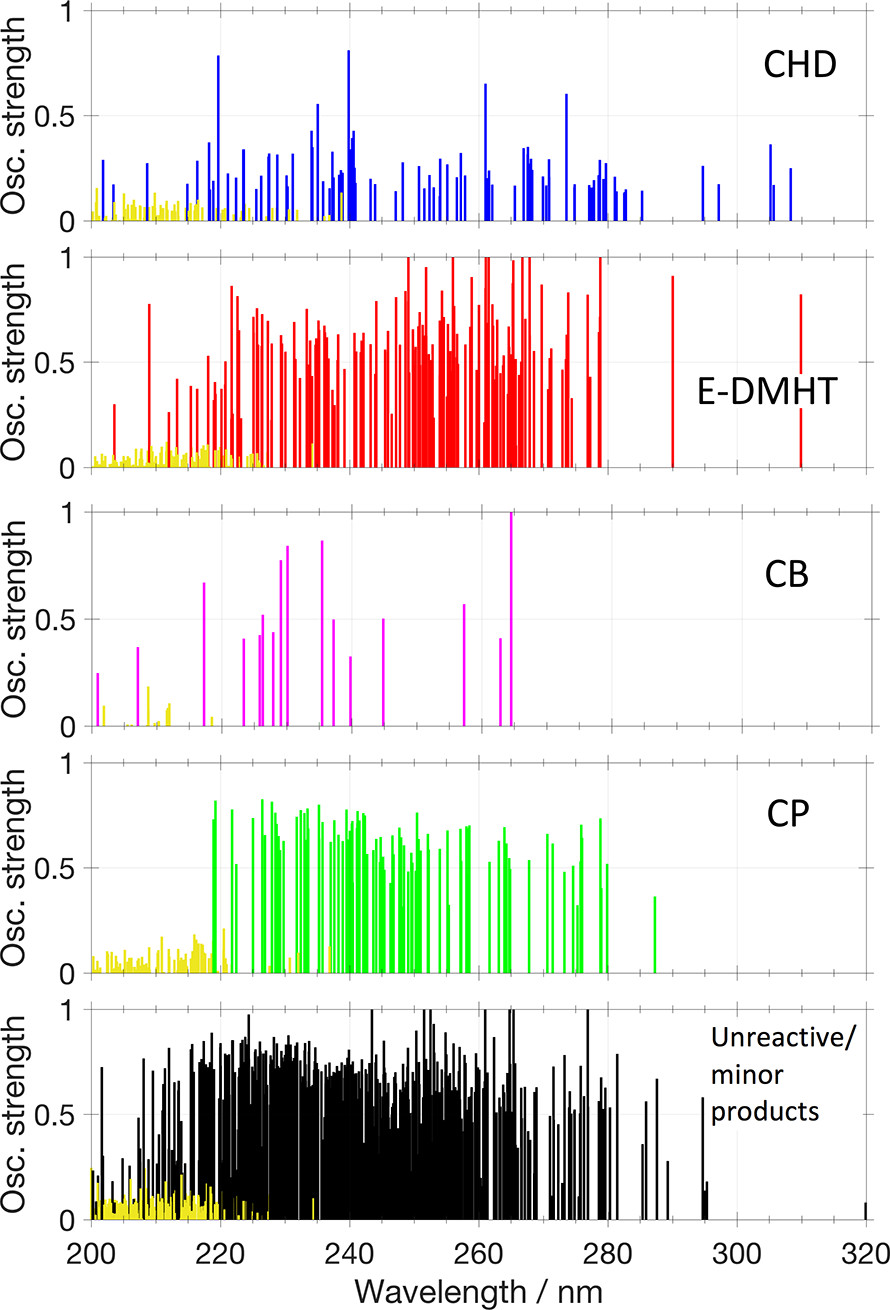}
    \caption{Distribution of the ADC(2) oscillator strengths of the first excited singlet state of the initial structures (blue, red, magenta, green, and black) of the Z-DMHT trajectories that form the indicated product. Yellow: oscillator strengths of the second excited singlet state. Reprinted with permission from Thompson and Tapavicza, J. Phys. Chem. Lett. 9, 4758–4764 (2018). Copyright 2019 American Chemical Society.}
    \label{fig:DMHT_osc_PQY}
\end{figure}

\begin{figure}
    \centering
    \includegraphics[scale=1.6]{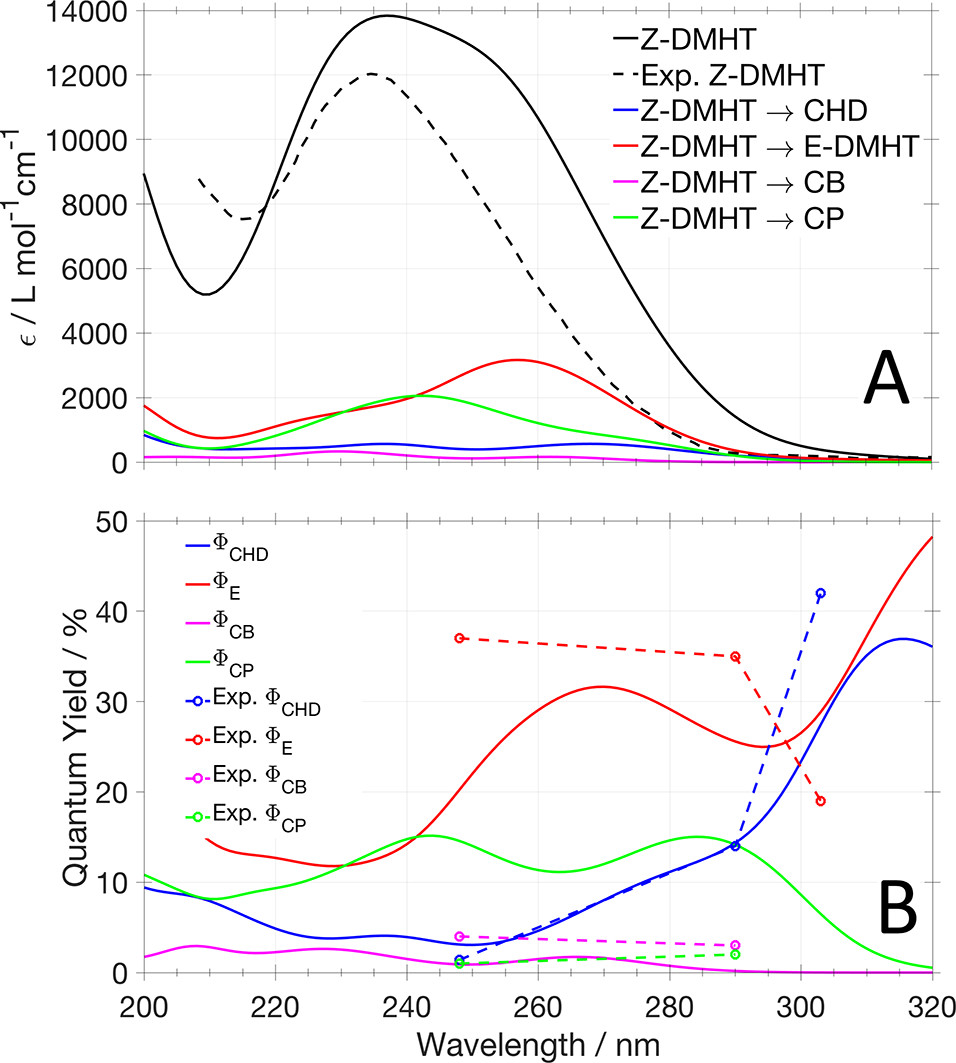}
    \caption{ (A) ADC(2) Absorption spectra of DMHT. The calculated absorption spectrum of DMHT averaged over all TDDFT-SH initial structures (solid) and experimentally measured spectrum\protect\cite{brouwer1987wavelength} (dashed) are shown in black. The spectra averaged over initial structures of the trajectories leading to the indicated products are given in color. (B) Calculated and experimentally measured\protect\cite{brouwer1987wavelength} (Exp.) wavelength-dependent product quantum yields of E-DMHT ($\Phi_E$), CHD ($\Phi_{CHD}$), CB ($\Phi_{CB}$), and CP ($\Phi_{CP}$). Reprinted with permission from Thompson and Tapavicza, J. Phys. Chem. Lett. 9, 4758–4764 (2018). Copyright 2019 American Chemical Society.}
    \label{fig:DMH_PQY}
\end{figure}

\begin{figure}
    \centering
    \includegraphics[scale=0.6]{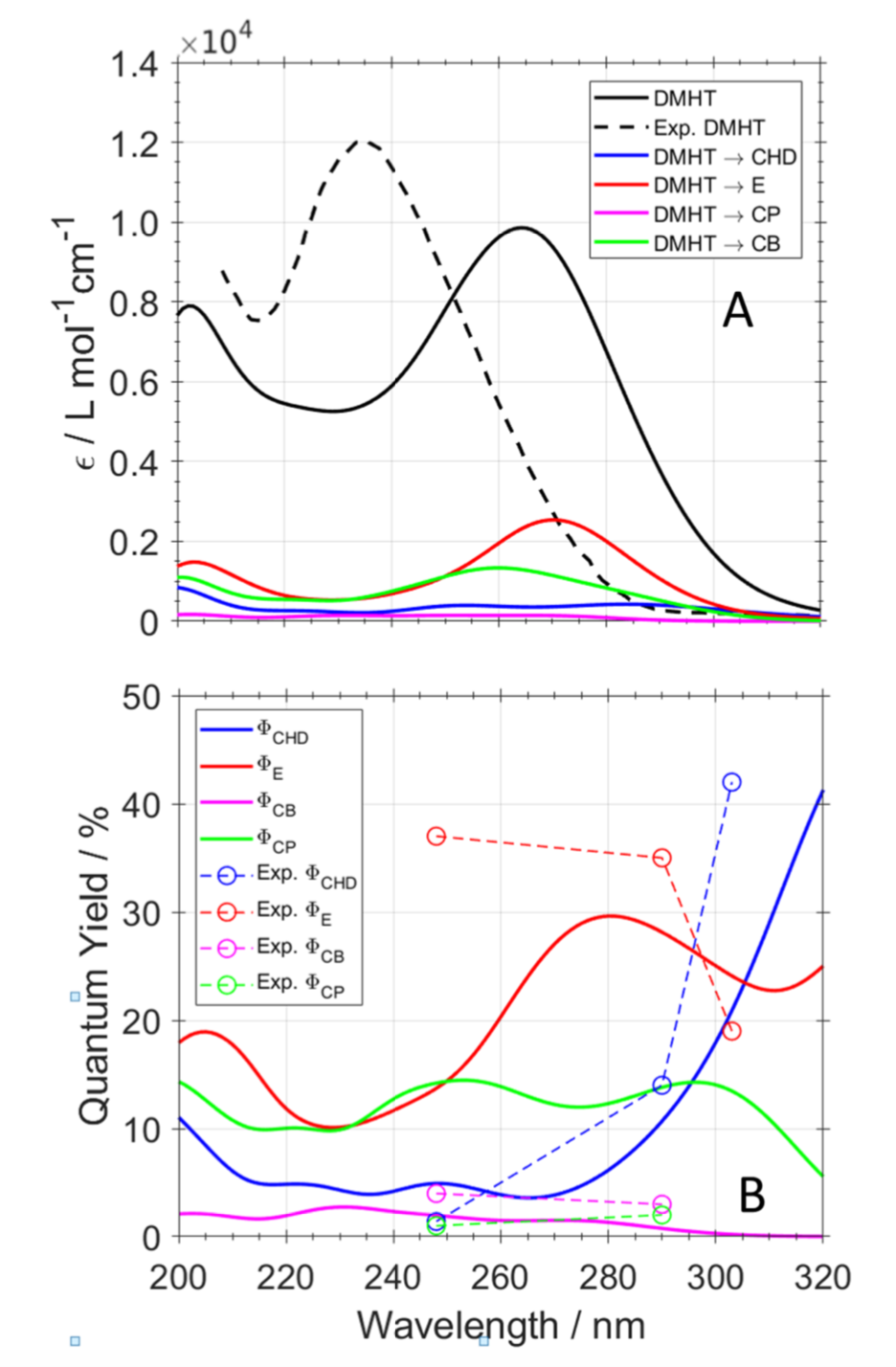}
    \caption{ (A) TDPBE0 Absorption spectra of DMHT. The calculated absorption spectrum of DMHT averaged over all TDDFT-SH initial structures (solid) and experimentally measured spectrum\protect\cite{brouwer1987wavelength} (dashed) are shown in black. The spectra averaged over initial structures of the trajectories leading to the indicated products are given in color. (B) Calculated and experimentally measured\protect\cite{brouwer1987wavelength} (Exp.) wavelength-dependent product quantum yields of E-DMHT ($\Phi_E$), CHD ($\Phi_{CHD}$), CB ($\Phi_{CB}$), and CP ($\Phi_{CP}$). Reprinted with permission from Thompson and Tapavicza, J. Phys. Chem. Lett. 9, 4758–4764 (2018). Copyright 2019 American Chemical Society.}
    \label{fig:DMH_PQY_TDDFT}
\end{figure}

In the previous sections we have seen the characteristic wavelength dependent photochemistry of 3Z-hexatriene and vitamin D derivatives. Now we will turn to the quantitative prediction of the wavelength dependent PQYs based on TDDFT-SH simulations and electronic spectra predictions.
We will focus on the calculations of the ring-closing and Z/E-isomerization PQYs of DMHT and Pre, because for these molecules experimentally determined PQYs are available in the literature for comparison \cite{brouwer1987wavelength,Dauben1991}.
For the quantitative prediction of the PQY, using Equation \ref{qy3}, we need to group the TDDFT-SH trajectories according to their product formed and compute the electronic absorption spectra averaged over the initial structures of each group.
For DMHT, the major reaction channels are ring-closing, Z/E-isomerization, formation of a cyclobutene derivative {\bf 11}, and formation of cyclopropan derivatives {\bf 8} and {\bf 9}, with branching ratios given in Figure \ref{fig:HT_channels}. We computed the excitation energies and oscillator strengths using TDDFT and the algebraic diagrammatic construction to second order (ADC(2)). Since experimental PQY measurements were carried out in n-pentane, the conductor like screening method (COSMO) \cite{Klamt1993} was used to mimic the solvent via a dielectric continuum model. Inspecting the distribution of ADC(2) oscillator strengths as a function of the S$_1$ and S$_2$ excitation energies (Figure \ref{fig:DMHT_osc_PQY}), we note that in the region between 240 and 320 nm, only oscillator strengths of S$_1$ are located, whereas S$_2$ oscillator strengths are located at wavelengths below 240 nm. Thus, the long wavelength photochemistry is mainly governed by S$_1$ photochemistry. For the ring-closing channel (CHD formation), several structures exhibit excitations in the long-wavelength region between 280 and 320 nm. This is consistent with our previous finding that only gZg conformers undergo ring-closing and that these conformers absorb at the red side of the absorption spectrum. For the Z/E-isomerization (E-DMHT formation) only two structures with excitation energies located in this region are present. To obtain the wavelength dependent PQY, we first converted oscillator strengths of the entire ensemble and for the different reaction channels into absorption spectra (Figure \ref{fig:DMH_PQY}, A for ADC(2) and Figure \ref{fig:DMH_PQY_TDDFT}, A for TDDFT). Dividing the spectrum of CHD formation by the total spectrum yields the wavelength dependent PQY for CHD formation.  Analogously, we obtain the PQYs of the other products. As expected from earlier considerations, we see an increase of CHD formation towards the red side of the spectrum, which is confirmed by the experimental measurements. For Z/E-isomerization, the experimental measurements show a decrease of the PQY with increasing wavelength. This is partially captured by the calculated PQY, which also shows a drop until 300 nm. However, above 300 nm we see an increase, but this increase is possibly an artifact due to the poor statistics in this wavelength region. It is probably caused by only one single trajectory, which shows the isolated excitation energy on the red side in the spectrum (Figure \ref{fig:DMHT_osc_PQY}).
Minor products, {\bf 11} and {\bf 9} show lower PQY than CHD and E-DMHT, in both, experiment and calculation. While the computed PQY for the cyclobutene product is in the same order of magnitude as in the experimental one, the cyclopropan derivative formation is overestimated by more than 10~\%. PQYs determined on the basis of TDDFT spectra agree similarly well with experimental results as the PQYs determined based on ADC(2) data. However, we see that the $\lambda_{max}$ value of the total TDDFT spectrum (Figure \ref{fig:DMH_PQY_TDDFT}, A) does not agree as well with experimental value as the $\lambda_{max}$ value of ADC(2) total spectrum  (Figure \ref{fig:DMH_PQY}, A). Due to the better agreement of the ADC(2) spectrum with experimental spectrum, one can also expect a more accurate prediction of the PQYs based on ADC(2) computed spectra.

\begin{figure}
    \centering
    \includegraphics{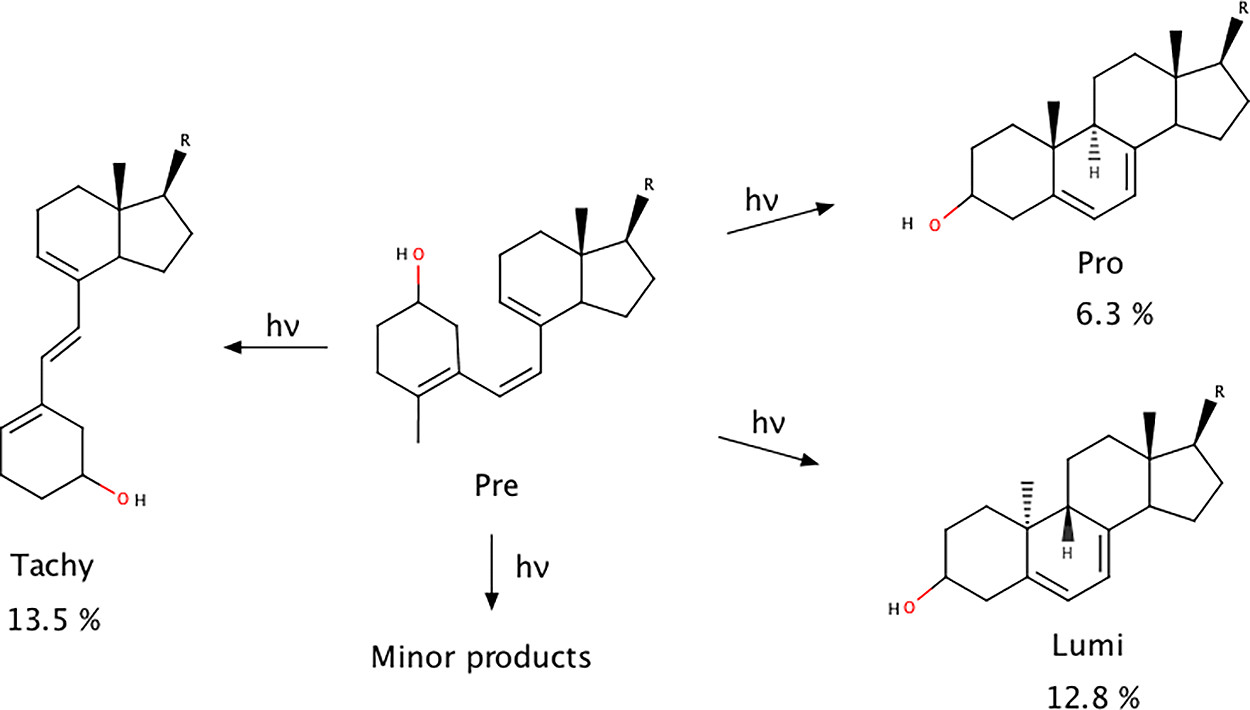}
    \caption{Major Reactions and Their Products Observed in TDDFT-SH Simulations of truncated Pre Z/E-isomerization, forming tachysterol, is shown on the left, and ring-closing reactions, forming lumisterol and provitamin D, is shown on the right. Raw branching ratios from simulations are indicated for the products. Reprinted with permission from Thompson and Tapavicza, J. Phys. Chem. Lett. 9, 4758–4764 (2018). Copyright 2019 American Chemical Society.}
    \label{fig:Pre_scheme}
\end{figure}

\begin{figure}
    \centering
    \includegraphics{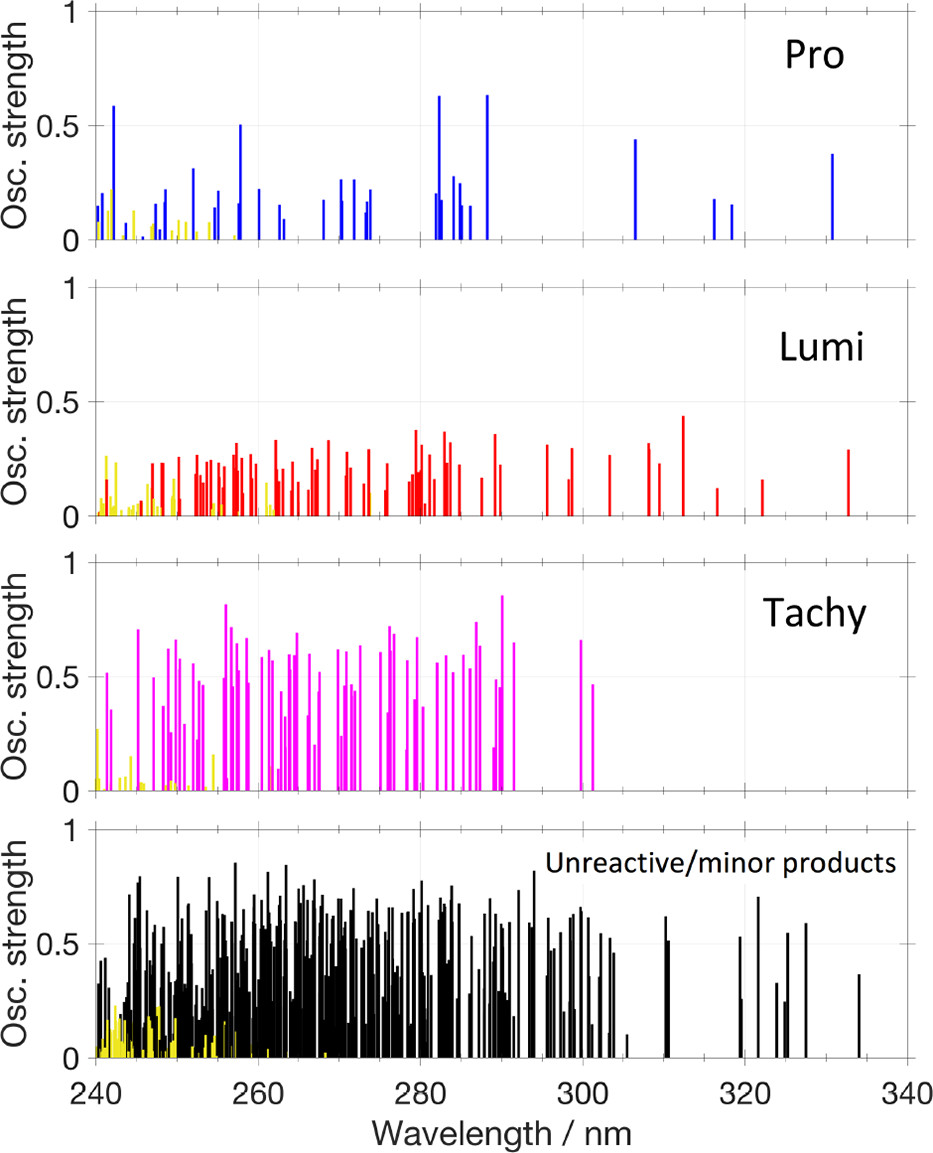}
    \caption{Distribution of the ADC(2) oscillator strengths of the first excited singlet state of the initial structures of the Pre trajectories that form the indicated product (blue, red, magenta, and black). Yellow: oscillator strengths of the second excited singlet state. Reprinted with permission from Thompson and Tapavicza, J. Phys. Chem. Lett. 9, 4758–4764 (2018). Copyright 2019 American Chemical Society.}
    \label{fig:PYQ_Pre}
\end{figure}

Turning to the calculation of the wavelength-dependent PQY for Pre, we first present the branching rations of the TDDFT-SH simulations (Figure \ref{fig:Pre_scheme}). In case of Pre, two distinct ring-closing products, Lumi and Pro are possible, originating from g(+)Zg(+) and g(-)Zg(-) conformers, respectively. Consistent with the dihedral plot (Figure \ref{fig:Pre_phi_exci}), there are only initial structures of trajectories forming Lumi and Pro above 300 nm on the red side of the spectrum (Figure \ref{fig:PYQ_Pre}).
In contrast, Tachy forming trajectories all originate from initial structures with S$_1$ excitation energies below 300 nm. We also note the larger oscillator strengths of Tachy forming trajectories, compared to Lumi and Pro forming initial structures. The large oscillator strengths are characteristic for the tZt structures, which are prone to undergo Z/E isomerization. Converting oscillator strengths to absorption spectra and computing the wavelength dependent PQYs, we note a good agreement between calculated and experimentally measured values for Lumi and Pro formation. Not only are calculations able to capture the increased ring-closing towards the red side of the spectrum, they also agree with experiments in the relative magnitudes between Lumi and Pro formation. This is a direct consequence of the higher amount of g(+)Zg(+) structures compared to g(-)Zg(-) structures, suggesting an accurate description of the conformational equilibrium by the REMD simulations.

\section{Conclusion and Outlook}
As we have seen from the comparison between non-adiabatic TSH simulations and experimental results, the simulations are able to explain conformationally controlled photochemistry in a variety of Z-hexatriene and vitamin D derivatives. An important ingredient to achieve this goal is to draw initial structures for the TSH trajectories from a Boltzmann ensemble of structures that contain the different conformational isomers with the correct statistical distribution. Provided an accurate potential energy function, either from an empirical force field or from {\it ab initio} methods is available, enhanced sampling methods such as the presented REMD can be used to obtain an accurate ensemble of structures. Here we have used DFT as underlying potential energy function for the initial sampling. However, common approximations to the exchange correlation functional are known to be problematic in the description of dispersion forces. This could affect the accuracy of the obtained ensemble of structures. Future benchmark studies with {\it ab initio} methods that are able to treat dispersion more accurately would be desirable and could possibly lead to a more accurate description of Boltzmann ensembles and in turn positively affect the accuracy of product distributions in TSH simulations.

In the calculation of wavelength dependent PQYs the description of the ground state ensemble of structures affects the accuracy of the PQYs in two ways: firstly, an accurate ensemble is necessary to obtain accurate absorption spectra and secondly, the correct statistical distribution is necessary to obtain correct branching rations of the different photochemical reaction channels. Regarding spectra calculations, we have also seen that highly accurate correlated excited methods such as ADC(2) are often superior to TDDFT. Equally, to obtain good comparability between calculated wavelength dependent PQYs with experimentally measured ones, correlated methods should be chosen for the spectra calculations. 
It remains the question regarding the accuracy of TDDFT for excited state dynamics. Some issues, such as erroneous description of conical intersections, charge transfer excitations, and double excitations have been mentioned here. These issues are an active field of research \cite{lee2019conical}.
While there have been some studies that employ correlated wavefunction based methods, such as CC2 \cite{Tapavicza2009,plasser2014surface} or CASPT2 \cite{barbatti2010relaxation,park2017fly} in TSH, often the computational resources available only allow TSH simulations based on TDDFT, therefore improvements of TDDFT regarding these issues
will increase the accuracy of TSH simulations in the future.

One point that was only briefly mentioned here, is the influence of the solvent the photochemical reactivity. While we used an electrostatic continuum model to account for electronic effects in the calculation of spectra, gauging the influence of solvent viscosity or viscosity of a membrane requires models that explicitly take into account the solvent molecules.
This requires a more sophisticated simulation setup. While efficient methods to include the solvent in these calculations have been developed (for instance hybrid quantum mechanical/molecular mechanics, QM/MM methods\cite{moret05,liu2017qm}), they have not yet been applied to compute wavelength dependent PQYs. 

Another aspect that possibly affect the outcome of TSH simulations is related to the initial conditions. The cases presented here are all based on initial structures and velocities obtained from MD simulations. Also here, a systematic comparison between TSH using MD-generated initial conditions and initial conditions from a Wigner distribution would be useful to direct future studies.

Besides, it would be interesting to investigate how the inclusion of nuclear quantum effects would affect the TSH simulation of the various photoinduced hydrogen transfer reactions. Methods such as the ring-polymer surface hopping dynamics\cite{shushkov2012ring} could be applied to investigate this question.

The large variety of different systems besides HT derivatives that exhibit conformationally controlled photochemistry are still up to study with non-adiabatic simulation techniques. New developments in more efficient excited state methods (e.g. tight binding TDDFT) will allow to also study larger systems and assist the development of photochemical switches and molecular motors in condensed systems. 

%FIGURES






\begin{figure}
    \centering
    \includegraphics[scale=1.2]{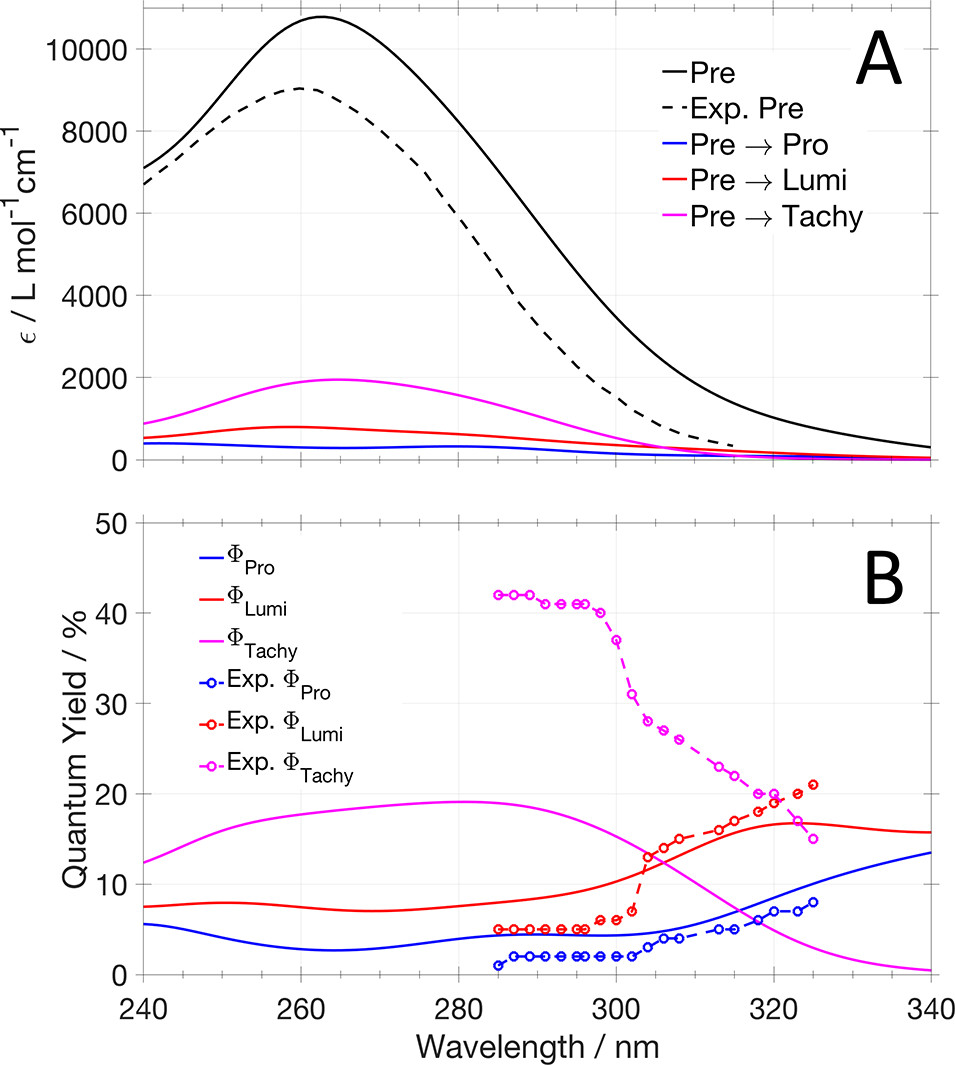}
    \caption{ (A) Absorption spectra of Pre. Calculated spectrum of all TDDFT-SH initial structures (solid) and experimentally measured spectrum\protect\cite{Dauben1991} (dashed) are shown in black. The spectra of the initial structures of the trajectories that lead to the indicated reactions are given in color. (B) Calculated and experimentally measured\protect\cite{Dauben1991} (Exp.) wavelength-dependent product quantum yields of Pre  and wavelength-dependent product quantum yields of Lumi ($\Phi_{Lumi}$), Pro ($\Phi_{Pro}$), and Tachy ($\Phi_{Tachy}$). Reprinted with permission from Thompson and Tapavicza, J. Phys. Chem. Lett. 9, 4758–4764 (2018). Copyright 2019 American Chemical Society.}
    \label{fig:my_label}
\end{figure}

\clearpage

\bibliography{papers}
\end{document}